\documentstyle[11pt,aaspp4,flushrt]{article}

\def\lya{Ly$\alpha$}
\def\N5{\ion{N}{5}~$\lambda$1240}
\def\siliox{\ion{Si/0}{2}~$\lambda\lambda$1400}
\def\sul5{\ion{S}{5}~$\lambda$1502}
\def\carbon4{\ion{C}{4}~$\lambda\lambda$1549}
\def\He2{\ion{He}{2}~$\lambda$1640}
\def\oxy3{\ion{O}{3}]~$\lambda\lambda$1663}
\def\carb3{\ion{C}{3}]~$\lambda$1909}
\def\car2{\ion{C}{2}]~$\lambda$2326}
\def\Ne4{[\ion{Ne}{4}]~$\lambda$2424}
\def\Mg2{\ion{Mg}{2}~$\lambda\lambda$2800}
\def\o2{[\ion{O}{2}]~$\lambda$3727}
\def\he2b{\ion{He}{2}~$\lambda$4686}
\def\O3{[\ion{O}{3}]~$\lambda\lambda$4959, 5007}

\def\spose#1{\hbox to 0pt{#1\hss}}

\def\kms{\ifmmode {\rm\,km\,s^{-1}}\else
    ${\rm\,km\,s^{-1}}$\fi}
\def\kmsMpc{\ifmmode {\rm\,km\,s^{-1}\,Mpc^{-1}}\else
    ${\rm\,km\,s^{-1}\,Mpc^{-1}}$\fi}

\def\msun{\ifmmode {\rm\,M_\odot}\else ${\rm\,M_\odot}$\fi}
\def\Msun{\ifmmode {\rm\,M_\odot}\else ${\rm\,M_\odot}$\fi}
\def\lsun{\ifmmode {\rm\,L_\odot}\else ${\rm\,L_\odot}$\fi}
\def\Lsun{\ifmmode {\rm\,L_\odot}\else ${\rm\,L_\odot}$\fi}
\def\rsun{\ifmmode {\rm\,R_\odot}\else ${\rm\,R_\odot}$\fi}
\def\Rsun{\ifmmode {\rm\,R_\odot}\else ${\rm\,R_\odot}$\fi}

\def\cm{{\rm\,cm}}
\def\cm3{\ifmmode {\rm\,cm^{-3}}\else ${\rm\,cm^{-3}}$\fi}

\def\ergps{\ifmmode {\rm\,erg\,s^{-1}}\else ${\rm\,erg\,s^{-1}}$\fi}
\def\ergpscm2{\ifmmode {\rm\,erg\,s^{-1}\,cm^{-2}}\else
    ${\rm\,erg\,s^{-1}\,cm^{-2}}$\fi}

\def\cf{{cf.,\,}}
\def\eg{{e.g.,\,}}
\def\deg{\ifmmode {^{\circ}}\else {$^\circ$}\fi}
\def\degr{\ifmmode {^{\circ}}\else {$^\circ$}\fi}
\def\degs{\ifmmode {^{\circ}}\else {$^\circ$}\fi}

\def\etal{{et al.~}}

\def\h3Mpc{h^{-3}{\rm Mpc}^3}
\def\Ho{\ifmmode {\rm\,H_\circ}\else ${\rm\,H_\circ}$\fi}
\def\hnot{\ifmmode {\rm\,H_\circ}\else ${\rm\,H_\circ}$\fi}
\def\h0{\ifmmode {\rm\,H_\circ}\else ${\rm\,H_\circ}$\fi}
\def\hnotunit{\ifmmode {\rm\,km\,s^{-1}\,Mpc^{-1}}\else
    ${\rm\,km\,s^{-1}\,Mpc^{-1}}$\fi}
\def\qnot{\ifmmode {\rm\,q_\circ}\else ${\rm q_\circ}$\fi}
\def\q0{\ifmmode {\rm\,q_\circ}\else ${\rm q_\circ}$\fi}
\def\ie{{i.e.}}
\def\mic{\ifmmode {\rm\,\mu m}\else ${\rm \mu m}$\fi}

\def\arcsec{\ifmmode {^{\prime\prime}}\else $^{\prime\prime}$\fi}
\def\asec{\ifmmode {^{\prime\prime}}\else $^{\prime\prime}$\fi}
\def\arcmin{\ifmmode {^{\prime}}\else $^{\prime}$\fi}
\def\amin{\ifmmode {^{\prime}}\else $^{\prime}$\fi}

\def\secper{\ifmmode \rlap.{^{s}}\else $\rlap{.}{^{s}} $\fi}
\def\minper{\ifmmode \rlap.{^{m}}\else $\rlap{.}{^m} $\fi}
\def\magper{\ifmmode \rlap.{^{m}}\else $\rlap{.}{^m} $\fi}
\def\farcs{\ifmmode \rlap.{^{\prime\prime}}\else
    $\rlap.{^{\prime\prime}}$\fi}
\def\arcsper{\ifmmode \rlap.{^{\prime\prime}}\else
    $\rlap.{^{\prime\prime}}$\fi}
\def\arcmper{\ifmmode \rlap.{^{\prime}}\else
    $\rlap.{^{\prime}}$\fi}
\def\spose#1{\hbox to 0pt{#1\hss}}
\def\simlt{\mathrel{\spose{\lower 3pt\hbox{$\mathchar"218$}}
     \raise 2.0pt\hbox{$\mathchar"13C$}}}
\def\simgt{\mathrel{\spose{\lower 3pt\hbox{$\mathchar"218$}}
     \raise 2.0pt\hbox{$\mathchar"13E$}}}

\def\araa{{ARA\&A}}
\def\aa{{A\&A}}
\def\aas{{A\&A}}

\def\aj{{AJ}}
\def\apj{{ApJ}}

\def\apjs{{ApJS}}

\def\mnras{{MNRAS}}
\def\nature{{Nature}}

\def\pasp{{PASP}}

\def\apjref#1;#2;#3;#4 {\par\pp#1, {#2}, #3, #4 \par}

\tighten

\slugcomment{\it to appear in The Astronomical Journal (March 1999)} 
 
\lefthead{Stern et al.}
\righthead{New Galaxies from the MG Catalog }

\begin{document}

\title{New High--Redshift Radio Galaxies from the MIT--Green Bank
Catalog\altaffilmark{1}}

\author{Daniel Stern}
\affil{Astronomy Department, University of California at Berkeley, CA 94720}
\affil{Electronic Mail: dan@copacabana.berkeley.edu}

\bigskip

\author{Arjun Dey\altaffilmark{2}}
\affil{Dept. of Physics and Astronomy, The Johns Hopkins University, 3400 N. Charles St., Baltimore, MD 21218}
\affil{Electronic Mail: dey@skysrv.pha.jhu.edu}

\bigskip

\author{Hyron Spinrad \& Leslie Maxfield }
\affil{Astronomy Department, University of California at Berkeley, CA 94720}
\affil{Electronic Mail: spinrad@bigz.berkeley.edu,lmm@astron.berkeley.edu}

\bigskip

\author{Mark Dickinson\altaffilmark{3}}
\affil{Dept. of Physics and Astronomy, The Johns Hopkins University, 3400 N.
Charles St., Baltimore, MD 21218}
\affil{Electronic Mail: med@stsci.edu}

\bigskip

\author{David Schlegel}
\affil{Department of Astrophysical Sciences, Princeton University, Princeton,
NJ 08544}
\affil{Electronic Mail: schlegel@astro.princeton.edu}

\bigskip

\author{Rosa A. Gonz\'alez }
\affil{Space Telescope Science Institute, Baltimore, MD 21218}
\affil{Electronic Mail: ragl@stsci.edu}

\bigskip
\bigskip
\bigskip
\bigskip
\bigskip
\bigskip
\bigskip
\bigskip

\altaffiltext{1}{Based on observations obtained at the W.M. Keck
Observatory, Lick Observatory, and the MDM Observatory.  The Keck
Observatory is operated as a scientific partnership among the
University of California, the California Institute of Technology, and
the National Aeronautics and Space Administration, and was made
possible by the generous financial support of the W.M. Keck Foundation.}

\altaffiltext{2}{Hubble Fellow.}
\altaffiltext{3}{Alan C. Davis Fellow, also at the Space Telescope Science
Institute.}

\eject

\begin{abstract}

We present optical identifications and redshifts for seventeen new
high--redshift radio sources.  Fifteen of these sources are radio
galaxies; the remaining two are high--redshift, steep--spectrum,
radio--loud quasars. These objects were discovered as part of an
ongoing study of compact ($\theta < 10$\arcsec), moderately steep
spectrum ($\alpha_{\rm 1.4~GHz}^{\rm 4.8~GHz} > 0.75,
S_\nu\propto\nu^{-\alpha}$) sources from the MIT -- Green Bank (MG)
radio catalog ($S_{\rm 5 GHz} \simgt 50$ mJy).  Spectra for the optical
counterparts were obtained at the W.M. Keck Telescopes and are among
the optically faintest radio galaxies thus far identified.  Redshifts
range between 0.3 and 3.6, with thirteen of the seventeen at redshifts
greater than 1.5.  Combining these new radio galaxies with two
published MG radio galaxy spectra, we synthesize a composite MG radio
galaxy spectrum and discuss the properties of these galaxies in
comparison to other, more powerful, radio galaxies at similar
redshifts.  We suggest a radio power---ionization state relation.

\end{abstract}

\keywords{cosmology: early universe --- galaxies: active --- galaxies: redshifts -- galaxies: evolution -- radio continuum: galaxies}

\eject

\section{Introduction}

The MIT -- Green Bank (MG) survey (Bennett \etal 1986; Lawrence \etal
1986) consists of 5974 radio sources in 1.87 steradians of sky in an
equatorial strip which have flux densities $\simgt 50$ mJy at 5 GHz.
Studies of higher--flux density radio surveys such as the 3CR (Spinrad
\& Djorgovski 1987), Molonglo (McCarthy \etal 1996), and B2/1 Jy of
Allington--Smith (1982) have yielded many interesting discoveries, such
as the radio--optical alignment of high--redshift radio galaxies
(Chambers \etal 1987; McCarthy \etal 1987), very high--redshift
galaxies (\eg Lacy \etal 1995; Spinrad, Dey, \& Graham 1995; Rawlings
\etal 1996), and rich clusters at high--redshift (\eg Dickinson
1996).  The MG survey probes the parameter space between high flux
density surveys such as the 3CR, and milli-- and micro--Jansky surveys
such as the Leiden Berkeley Deep Survey and its extensions (Neuschaefer
\& Windhorst 1995).  The MG range of radio flux density is also
considered by the B3 survey (Vigotti \etal 1989) and recent Cambridge
surveys (\eg 6C --- Hales, Baldwin, \& Warner 1993), but the MG survey
was conducted at a higher frequency than these other surveys.
Selecting an intermediate power sample allows us to study various
properties as a function of radio power.  In particular, a lower radio
power suggests these galaxies may, on average, be less dominated by
their active nuclei, and may thus be more representative of giant
ellipticals at large redshift (\cf Dunlop \etal 1996; Spinrad \etal
1997; Dey \etal 1998).  Fainter flux densities also suggest that a
fraction of our sample may be at very high redshift, and indeed the
median redshift of radio galaxies in the MG observed to
this point is $z_{\rm med} \sim 1.1$ compared to $z_{\rm med} = 0.27$
for the 3CR (McCarthy 1993). 

Over the past several years we have been pursuing the optical
identification of a subset of 218 radio sources systematically selected
from the MG catalog.  We restrict our sample to radio sources of small
angular size ($\theta \leq 10 \arcsec$), simple (unresolved, double, or
triple) radio morphology, and moderately steep radio spectral index
($\alpha_{\rm 1.4~GHz}^{\rm 4.8~GHz} > 0.75,
S_\nu\propto\nu^{-\alpha}$).  The size and morphological criteria
primarily select against objects at low redshift.  The angular size
limit should also bias the sample to larger lookback times when the
higher density of the intergalactic medium is more effective at
confining radio lobes.  The spectral index criterion eliminates
low--redshift interlopers and selects against flat--spectrum quasars
and in favor of radio galaxies, but is only moderately restrictive
amongst the radio galaxies themselves (Blumenthal \& Miley 1979).
Resulting identifications are often very distant objects and thus serve
as useful probes for the study of galaxy formation and evolution.  In
combination with other complete samples, such as the 3CR, optical and
near--IR properties of MG sources can be used in correlative studies
which span a range of radio power at any given redshift.  Similar
studies suggest that weak radio sources ($S_{\rm 1.4 GHz} < 50$ mJy)
have weaker emission lines and the alignment effect becomes less
pronounced with lower radio flux density samples (\eg Rawlings \&
Saunders 1991; Dunlop \& Peacock 1993; Eales \& Rawlings 1993; Thompson
\etal 1994).  McCarthy (1993) has a recently presented a comprehensive
review on the optical and associated properties of radio galaxies.

Preliminary results from our survey are described in Spinrad \etal
(1993), Dey, Spinrad, \& Dickinson (1995), and Stern \etal (1996;
1997).  Currently, we have optical identifications for $\sim 85$\% of
our sample and spectroscopic redshifts for $\simgt 60$\% of the
sample.    In this paper we report on seventeen new identifications and
redshifts of MG sources.  The spectra were obtained at the W.M. Keck
Telescopes and are representative of some of the fainter radio
identifications thus far.  Typical $R$ magnitudes are $23 - 24.$ Many
of these sources are in the so--called `redshift desert,' $1.4 \simlt z
\simlt 2.0,$ for which no strong features (\ie, Ly$\alpha$, \o2, or
H$\alpha$) are redshifted into the optical window, thus making redshift
determinations challenging.  Our MG sample currently includes 13
galaxies with $z>2$; we report on 4 such galaxies here.

The paper is organized as follows:  in \S2 we present the observations
and data reductions, followed by a discussion of individual systems in
\S3.  We construct a composite spectrum and compare it with models and
other composite spectra of active galaxies in \S4.  A concluding
discussion comprises \S5.  Throughout this paper we adopt $H_0 = 50\,
h_{50}\, \kmsMpc,\, q_0 = 0.5,\,$ and $\Lambda = 0$ unless otherwise
noted.

\section{Observations and Data Reduction}
\subsection{Imaging and Optical Identifications}

Preliminary imaging observations of all sources were made over the
course of several years (1988 $-$ 1998) at the Lick Observatory 3m
Shane Telescope on Mount Hamilton and the MDM 2.4m Hiltner Telescope at
Kitt Peak.  The Kast Double Spectrograph of Lick Observatory (Miller \&
Stone 1994) employs UV--flooded Reticon 1200x400 CCDs with 27$\mu$m
pixels, corresponding to a plate scale of 0\farcs78 pix$^{-1}.$ Typical
seeing for the Lick imaging was 1\farcs2 -- 1\farcs8.  The Charlotte
camera of MDM employs a Tek 1024$^2$ CCD detector with a plate scale of
0\farcs275 pix$^{-1}.$ Typical seeing for these images was 1\farcs0 --
1\farcs5.

A subsample of seven of the seventeen sources were later reimaged with
the Low Resolution Imaging Spectrometer (LRIS; Oke \etal 1995) at the
10m W.M. Keck Telescopes on Mauna Kea between 1994 and 1998.  The
detector is a Tek 2048$^2$ CCD with 24$\mu$m pixels.  Due to a dewar
change in July 1996, the pixel scale changed from 0\farcs214 pix$^{-1}$
to 0\farcs212 pix$^{-1}$ at that time.  The typical seeing was 0\farcs8
-- 1\farcs0.  A journal of our observations is provided in
Table~\ref{obsns}.

The images were corrected for overscan bias, flattened using a median
sky flat, coadded, and calibrated using observations of standard
stars.  Utilizing astrometry provided by B. Burke and S. Conner, we
obtained optical identifications for our candidates based upon the
radio coordinates.  Typically the optical morphologies are small,
faint, and round.  We present finding charts in Fig.~\ref{plate}
(Plates 1 -- 3).  Where available, we present Keck images; the
remaining images were obtained at Lick and the MDM as indicated in
Table~\ref{obsns}.  Astrometry for each identification is tabulated in
Table~\ref{astrom} with the coordinates of the offset star provided.
The estimated uncertainty in the optical positions is 0\farcs3.
Optical and radio properties of the sources are in Table~\ref{prop},
where the radio flux densities and spectral indices are from Bennett
\etal (1986) and Lawrence \etal (1986).  Radio luminosities have been
calculated at a rest--frame frequency of 4.8~GHz, implementing the
spectral index derived from 1.4 and 4.8 GHz observations, a Hubble
constant of $H_0 = 50\, h_{50}\, \kmsMpc,\, q_0 = 0,\,$ and $\Lambda
= 0$.  In Table~\ref{radflux} we include more recent 4.85 GHz flux
densities for the sample.  Many sources may vary at the 10\% level over
the 10 year baseline, though MG~2058+0542 apparently varied
considerably more over the same period.  We therefore indicate its
spectral index as uncertain in Table~\ref{prop}.  Magnitudes, measured
in site--dependent apertures as described in Table~\ref{prop}, are from
our CCD photometry and have errors of $\sim 0.1$ to $0.3$ magnitudes.
Observations obtained at Lick Observatory employed the Spinrad
night--sky filter and yield red magnitudes in the $R_S$ system, which
is related to commonly used photometric systems by Djorgovski (1985)
\footnotemark[1].
Typical optical magnitudes are $R \sim 23 - 24,$ probing the fainter
envelope of our MG survey.

\footnotetext[1]{The transformation from Spinrad night--sky $R_S$ into
Johnson $VR$ is $(R_S - R) = -0.004 - 0.072(V-R) + 0.073(V-R)^2.$}

\subsection{LRIS Spectroscopy}

We obtained spectroscopic observations with LRIS between 1994 and
1997.  We generally used a 300~$\ell$/mm grating (blazed at 5000~\AA)
to sample a wavelength range $\lambda\lambda$4000$-$9000~\AA, and a 1
arcsecond wide slit which yields an effective resolution FWHM of
$\approx$ 10~\AA.  The read noise and gain were typically 8.0$e^-$ and
1.6$e^-$ adu$^{-1}$ respectively.  However, during the 1994 March run,
the LRIS CCD was affected by pattern noise which mimicked extremely
high read noise (25$e^-$ and 65$e^-$ on the blue and red sides),
noticeably degrading the spectrum of MG~1251+1104.  Observations were
typically done at an airmass $\simlt 1.05$. Hence, although the slit
position angle was often selected to trace the radio morphology or to
include nearby optical sources, we believe our relative
spectrophotometry to be accurate for the wavelengths considered (\cf
Filippenko 1982).  The sources attempted spectroscopically at Keck were
primarily selected from the fainter ($R \simgt 23$) identifications in
the MG survey, many having been unsuccessful spectroscopic targets at
Lick Observatory.  Exceptions are the relatively bright sources
MG~0422+0816 ($R \sim 20$) and MG~0511+0143 ($R \sim 22$) which were
first observed spectroscopically as a backup project at Keck
Observatory on an inclement night.  These observations utilized a
600~$\ell$/mm grating (blazed at 5000~\AA).  The bright quasar
MG~2041+1854 ($R \sim 20$) was observed at Keck during twilight on a
cirrusy night, while the relatively bright ($R_S \sim 21$) sources
MG~0148+1028 and MG~0308+0720 were first surveyed spectroscopically as
backup targets on 23 December 1997 UT.

For each observation, we offset the telescope from the reference star
listed in Table~\ref{astrom}.  The data were corrected for overscan
bias, flat--fielded using internal lamps taken after each observation,
flux calibrated using observations of spectrophotometric standard stars
(Massey \etal 1988; Massey \& Gronwall 1990), and corrected for
Galactic reddening using the Burstein \& Heiles (1982) maps (see
Table~\ref{prop}) with the Cardelli, Clayton, \& Mathis (1989)
extinction curve.  One--dimensional spectra were extracted with a
typical aperture size of 1\farcs4.  The standard stars were observed
both with and without an OG570 filter in order to correct for the
second order light contaminating the wavelength region $\lambda >
7500~$\AA.  A journal of the observations is provided in
Table~\ref{obsns} and individual spectra are presented in
Fig.~\ref{spectra} with the prominent features indicated.  In
Table~\ref{lines} we list the line flux densities measured for each
source.  For completeness, we include the only two published spectra of
MG radio galaxies (MG~1019+0534 --- Dey, Spinrad, \& Dickinson 1995;
MG~2144+1928 --- Maxfield \etal 1998) in Fig.~\ref{spectra} and
Table~\ref{lines}. 

\section{Notes on Individual Sources}

{\em MG~0018+0940, at z=1.586,} exhibits a low ionization spectrum:
unlike more powerful radio sources, the \carb3 line is
stronger than the \carbon4 doublet.  \Mg2 is clearly resolved.  The
optical ($R$) image of MG~0018+0940 has an extended, chain--like
morphology oriented at a position angle of 137\deg\ in the optical.
The radio source is unresolved in 4.85 GHz observations with the
B--array of the VLA ($\sim$ 1\farcs2 beam; Lawrence \etal 1986).

{\em MG~0046+1102, at z=1.813,} also exhibits a low ionization
spectrum, with similar carbon and magnesium line strengths to
MG~0018+0940.  \Mg2 is again a strong and clearly resolved feature.
The optical identification is extended with a position angle of
130\deg\ in $R$, while the source is unresolved in 4.85 GHz
observations obtained with the B--array configuration of the VLA.

{\em MG~0122+1923, at z=1.595,} also exhibits a low ionization spectrum,
with carbon line strengths similar to the prior two radio galaxies.
The optical identification contains a compact core and diffuse
emission to the North oriented at a position angle of 41\deg.

{\em MG~0148+1028, at z=2.845,} is an intriguing source showing both
the bright emission lines typical of high--redshift radio galaxies and
the narrow absorption lines that characterize high--redshift,
star--forming galaxies (\cf Steidel \etal 1996, Lowenthal \etal
1997).  These lines, indicated with vertical dashed lines in Fig.~2,
refer to \ion{Si}{2}~$\lambda$1260, \ion{O}{1}~$\lambda$1302,
\ion{C}{2}~$\lambda$1335, \ion{Si}{4}~$\lambda\lambda$1394, 1403,
\ion{S}{5}~$\lambda$1502, \ion{Si}{2}~$\lambda$1526,
\ion{Fe}{2}~$\lambda$1608, and \ion{Al}{2}~$\lambda$1670 (\eg Spinrad
\etal 1998).  The emission lines of this source are slightly broader
than in those sources considered above, with FWHM $\sim$ 1000 km
s$^{-1}$ --- a value commonly seen in quasars, though the low
\carbon4/\He2 ratio leads us to classify this source as a broad--lined
radio galaxy.  A detailed analysis of MG~0148+1028 is deferred to
future publication; note, however, that \sul5, an unambiguous signature
of starlight (\eg Dey \etal 1997), is visible in absorption.  The
finding chart for this galaxy derives from a short (60$s$) exposure
obtained through LRIS without a filter to identify the field; this
image is superior to the 1260$s$ Lick image.  Slit spectroscopy was
done at a position angle of 138\deg\ so that galaxies b and c (see
Fig.~1) would also be observed.  Neither is associated with
MG~0148+1028; galaxy b has \o2 and \O3 in emission at $z=0.588,$ while
galaxy c has \o2 in emission with \Mg2 in absorption at $z=1.297.$
The optical identification is marginally resolved:  FWHM$_{\rm MG} \approx
1\farcs0$, while the seeing was $\approx 0\farcs9$.

{\em MG~0308+0720, at z=2.975,} is a radio--loud quasar as evidenced by
the broad emission lines (\eg FWHM(\ion{C}{4}) $\sim$ 5600 km
s$^{-1}$) and unresolved morphology. The Ly$\alpha$--\N5 emission
complex is unusual, with narrow Ly$\alpha$ superimposed upon a broad
\N5 emission line.  Recent statistical investigations of broad UV lines
in luminous (radio--quiet) quasars have suggested that the traditional
broad--line region (BLR) consists of two components --- a low--density,
intermediate--line region (ILR; FWHM $\sim$ 2000 km s$^{-1}$) and a
higher density, very broad line region (VBLR; FWHM $\simgt$ 7000 km
s$^{-1}$) blueshifted by $\simgt$ 1000 km s$^{-1}$ with respect to the
ILR (\eg Brotherton \etal 1994).  Differences in the relative
contributions of these two components can explain much of the observed
variation in quasar broad--line profiles.  For instance, Ly$\alpha$
will come predominantly from the ILR, while higher ionization lines
such as \N5, \ion{Si}{4}~$\lambda\lambda$1394, 1403, and \carbon4 are
dominated by VBLR emission.  In this simple model, MG~0308+0720 is a
quasar whose broad--lines are dominated by the VBLR; indeed, the
Ly$\alpha$ spectral profile looks similar to that of the composite VBLR
presented in Fig.~2a of Brotherton \etal (1994).  Alternatively, we may
be seeing a dusty quasar at high redshift.  The Ly$\alpha$--\N5 complex
is dominated by broad \N5 emission.  Subtracting a Gaussian fit to the
\N5 line, we find the residual Ly$\alpha$ to be extremely weak:  the
Ly$\alpha$/\carbon4 ratio is $\sim 0.25,$ compared to a mean value of
$1.2 - 2.2$ in composite quasar spectra (Boyle 1990; Francis \etal
1991).  The extremely red continuum is in concordance with this
dust--obscuration interpretation.  Three \Mg2 absorption systems at
redshifts $z=1.391, 1.607,$ and $1.814$ are also evident in the
spectrum.

{\em MG~0311+1532, at z=1.989,} exhibits a slightly higher ionization
spectrum in comparison to the above radio galaxies.  \car2 is barely
apparent as a faint broad bump near 7000~\AA.  The red continuum of
this source, and particularly the abrupt feature at 7150~\AA\, are
likely artifacts due to scattered light from a bright star 3\farcs5 to
the NE.  The optical identification consists of low surface brightness
emission oriented at a position angle of 135\deg, while the 4.8 GHz
morphology is extended by 5\farcs1 at a position angle of 161\deg.

{\em MG~0422+0816, at z=0.289,} was observed spectroscopically at Keck
as a backup object on a night of poor seeing and thick cirrus, thus
making our primary faint targets untenable.  The relatively bright
optical flux ($R \sim 20$) and unfluxed spectrum are the byproducts of
these meteorological conditions which prevented observations of our
primary fainter targets.  The \O3 nebular lines are the most prominent
feature of the spectrum.  The optical counterpart consists of a
marginally resolved core with diffuse emission to the south.

{\em MG~0511+0143, at z=0.596,} was also observed as a back--up target
on an inclement night.  This galaxy has a weak--lined spectrum with \o2
of low equivalent width and a prominent 4000~\AA\, break (labeled with
an arrow in Fig.~\ref{spectra}), indicative of an old stellar
population.  MG~0511+0143 was also imaged in the J and K' bands at Lick
Observatory on cirrusy nights.  Comparing the multiband images suggests
the radio galaxy may be part of a cluster at $z \sim 0.6.$ The
vignetting of the finding chart is due to the small field--of--view of
the Kast imaging spectrometer on the Lick 3m.  The optical
identification is oriented at a position angle of 48\deg, while
the 4.8 GHz morphology is extended by 3\farcs9 at a position angle
of 171\deg.

{\em MG~1019+0534, at z=2.765,} has an unusual spectrum which is
previously reported in Dey, Spinrad, \& Dickinson (1995).  Ly$\alpha$,
usually the strongest line in high--redshift radio galaxies, is very
weak.  This is likely due to dust attenuating the \lya\, emission,
implying dust formation at early epochs for this system.

{\em MG~1142+1338, at z=1.279,} is a weak--lined radio galaxy with
strong \o2 and high ionization [\ion{Ne}{5}]~$\lambda$3426.  \carb3 is
barely visible as a broad feature while neither \car2 nor \Mg2 are
visible.   The optical counterpart is compact.

{\em MG~1251+1104, at z=2.322,} was observed during the 1994 March run
when the LRIS CCD was compromised by high pattern noise.  Ly$\alpha$ is
the dominant feature of the spectrum.  The continuum slope is
untrustworthy and a relic of instrumental problems, and lead to
uncertain equivalent widths in Table~5.  The optical morphology is
diffuse and symmetric.

{\em MG~1401+0921, at z=2.093,} exhibits a moderate ionization spectrum
radio galaxy with strong \carb3 and weak \car2.  The optical
counterpart is slightly elongated with a shell structure to the NW,
while the radio morphology has a classic double structure of separation
3\farcs5 at a position angle of 138\deg.  A similar asymmetric optical
structure is observed in $V$--band images of 4C~23.56 (Knopp \&
Chambers 1997).  Deep multi--band images lead them to interpret the
morphology of 4C~23.56 as deriving from a dusty galaxy illuminated by a
beam from an active nucleus which is scattered into our line of sight.

{\em MG~2037$-$0011, at z=1.512,} is a weak--lined source showing
several faint emission lines.  Most strikingly, the \carb3 line has a
narrow component superimposed upon a broad component.  A comparison
with the similar redshift radio galaxy MG~0018+0940 is interesting ---
in particular, note that \Mg2 is not seen in the spectrum of
MG~2037$-$0011.  

{\em MG~2041+1854, at z=3.056,} is a radio--loud quasar.  The spectral
index of this source is very close to the lower limit of our sample,
which was imposed to separate the typically flat--spectrum quasars from
the typically steeper-spectrum radio galaxies.  The relatively bright
($R \sim 20$) source is near the Galactic plane.  Therefore, we took
advantage of the good seeing conditions at Mauna Kea to take a short
exposure spectrum of this source during twilight on a cirrusy night,
though, consequently, the spectral fluxes are uncertain.  The best
image available for this source is from a 5$s$ exposure through the
guide camera at Keck (Fig.~\ref{plate}; 0\farcs275 pix$^{-1}$).  The
field of view is smaller than for the other objects in Fig.~\ref{plate}
and the slit is visible; nevertheless, it is vastly superior to the
poor--seeing Lick 3m image.  The spectrum shows the typical broad lines
of a quasar, with intervening \Mg2 absorption systems apparent at
$z=0.875, 1.469, 1.914,$ and $1.946.$  This source is interesting in
terms of being a moderately steep--spectrum, radio--loud quasar at $z
\simgt 3.$  Few such sources are in the literature and may be
interesting in terms of unified models of extragalactic radio sources.

{\em MG~2058+0542, at z=1.331,} is typical of the sources discussed
here.  Narrow lines are visible with a faint red continuum, possibly
suggestive of star light.

{\em MG~2109+0326, at z=1.636,} exhibits a slightly higher ionization
spectrum than MG~2058+0542, or possibly reflects multiple ionization
states, as indicated by the relatively strong \Ne4 emission line.  An
emission line galaxy at $z=0.790$ was also serendipitously discovered
on the Keck spectrogram of this target.

{\em MG~2121+1839, at z=1.861,} shows neither the low--ionization \car2
nor the high ionization \Ne4 lines; the major features are \carbon4 and
\carb3.  Ly$\alpha$ has been detected for this object on the blue
camera of the Kast Double Spectrograph of Lick Observatory.  Keck/NIRC
$K$ band images of this target are presented in van Breugel \etal
(1998) and reveal a relatively smooth morphology which is not aligned
with the radio axis ($PA_{\rm 4.8 GHz} = 145\deg$, extended by
6\farcs3).  Serendipitous emission line galaxies at $z=0.353$ and
$z=0.859$ were also discovered on the LRIS spectrogram.

{\em MG~2144+1928, at z=3.592,} is the highest redshift radio galaxy
yet discovered in our MG sample.  Its redshift was first measured at
Lick Observatory and is mentioned in Spinrad \etal (1993).
MG~2144+1928 is an aligned radio galaxy with multiple components
showing interesting velocity structure.  Detailed discussion and
analysis of the optical spectrum, as well as astrometry and a finding
chart, is presented in Maxfield \etal (1998).  Near--infrared
observations of this galaxy are also presented in Armus \etal (1997)
and van Breugel \etal (1998).  Both the emission--line free $K'$ image
of van Breugel \etal (1998) and the narrow--band 2.3$\mu$m image of
Armus \etal (1997), selected to target the redshifted \O3 doublet, show
the host galaxy to be extended along the radio axis ($PA_{\rm 4.8 GHz}
= 177\deg$, extended by 8\farcs5).

{\em MG~2308+0336, at z=2.457,} is a high--redshift radio galaxy
exhibiting a spectrogram indicative of mixed ionization state.  Note
that \car2, the strongest of the carbon lines for this object, is quite
broad.  Ly$\alpha$ was also detected at Lick Observatory with the Kast
Double Spectrograph.  The optical counterpart consists of higher
surface brightness core oriented at a position angle of 137\deg, with
fainter emission evident to the south.  The 4.8 GHz radio axis is
aligned with the fainter emission ($PA_{\rm 4.8 GHz} = 175\deg$) and is
extended by 3\arcsec.

\section{Emission Line Properties and Ionization State }

In Fig.~\ref{compfig} we present a composite Keck/MG radio galaxy
spectrum constructed from eleven of the spectra presented here,
omitting the unfluxed spectra, the quasars, the extremely weak--lined
MG~1142+1338, and MG~1251+1104 whose spectrum is heavily affected by
pattern noise.  We also include MG~1019+0534 and MG~2144+1928, the only
previously published Keck spectra of MG radio galaxies (Dey, Spinrad,
\& Dickinson 1995; Maxfield \etal 1998), to maintain the integrity of
the selection criterion:  namely, we combine all Keck spectra of MG
radio galaxies that we have amassed to date.  This selection criteria
ensures that we use only the highest signal--to--noise ratio data.  The
composite was constructed by shifting the individual spectra into their
rest frame, rebinning to a common linear wavelength scale with 2
\AA\ pix$^{-1}$ resolution, and scaling the spectra by the flux of the
\carb3 emission line.  We select the \carb3 feature since it is present
in most of the spectra considered herein --- in objects at both high
and low redshift, as well as objects exhibiting high and low ionization
spectra.  We justify this algorithm in that our primary objective is to
study the emission line properties of the MG sample; normalizing the
spectra over a longer wavelength (line--free) range would give undue
weight to the galaxies with the lowest continua.  The current algorithm
biases the continuum to the weakest--lined sources.  Accordingly, we
deemphasize conclusions drawn regarding the composite continuum.  The
red ends of the spectra were then trimmed to minimize contamination
from the telluric OH features, and, in the case of MG~0311+1532, the
effects of a nearby red star.  The resultant vectors were then
averaged.  Only the spectrum of MG~2144+1928 did not overlap with the
\carb3 emission line.  Instead, we normalized its spectrum to a
preliminary composite comprising the other 12 galaxies using the
\carbon4 doublet.  The final spectrum is comparable to the composite
radio galaxy spectrum constructed by McCarthy \& Lawrence (1998;
hereinafter McL98) primarily from 3C and MRC/1 Jy sources, which we
also present in Fig.~\ref{compfig}.

The emission line properties of the composite MG spectrum ($\lambda
\lambda 1100 - 2900$ \AA) are presented in Table~\ref{composite}.  We
note that the emission line widths have not been corrected for the
resolution of the spectrograph.  For typical line widths of 1100 km
s$^{-1}$ observed at 7500 \AA, the deconvolution correction is
$\approx$ 75 km s$^{-1}$.  Comparisons to the emission line strengths
determined from composite spectra of higher flux density radio galaxies
(McL98), QSOs (Boyle 1990; Francis \etal 1991), Seyfert II nuclei
(Ferland \& Osterbrock 1986), the LINER nucleus of NGC~4579 (Barth
\etal 1996), and models are presented in Table~\ref{compare}.  All line
profiles in the MG composite were fit with a simple Gaussian.  This
simple prescription overestimates the width of the doublet lines, most
drastically for \Mg2 whose rest--frame separation is 7.2~\AA.  The
typical FWHM for the remaining lines is 900 $-$ 1100 km sec$^{-1},$
lower than that found by McL98 for their composite radio galaxy
spectrum.  Unlike McL98, we do not find that the FWHMs increase towards
shorter wavelengths.  However, the constituent galaxies in the McL98
composite stem from a much larger redshift range ($0.16 < z < 3.13$),
allowing a longer baseline composite spectrum ($\lambda\lambda 800 -
5500$~\AA).  No strong trend is evident in their composite for the
restricted wavelength range where both composites overlap.  McL98 note
that redshift correlates with \o2 and \O3 line width in 3CR galaxies,
suggesting that the wavelength--line width
trend in McL98 likely reflects a redshift--line width correlation, in
the sense that higher redshift radio galaxies have broader emission
lines.  This is unlikely to be a selection effect, as the 3CR is nearly
completely identified.


We measure an average restframe equivalent width of 300~\AA\ for the
Ly$\alpha$ emission line in the Keck/MG radio galaxy sample.  For
comparison, from a sample of 28 radio galaxies with $1.7 < z < 3.5,$
similar to the redshift range in the current sample, McCarthy (1993)
finds a mean rest--frame $W_\lambda({\rm Ly}\alpha)$ of $295 \pm
188$~\AA.  Radio--quiet, Lyman--break galaxies with Ly$\alpha$ in
emission have a typical $W_\lambda({\rm Ly}\alpha)$ of $3 -
20$~\AA\ (Steidel \etal 1996).  McL98 find that their composite radio
galaxy spectrum has very strong Ly$\alpha$ relative to \N5, \carbon4,
and \carb3, as compared to other AGN.  The MG composite, however, has
much weaker Ly$\alpha$, with line ratios more typical of the other AGN
in Table~\ref{compare}.  Most likely the radio galaxy discrepancies
stem from the small number of Ly$\alpha$ emitters in the composite
spectra.  The 3C/MRC composite Ly$\alpha$ comprises 7 galaxies, while
the MG composite Ly$\alpha$ comprises only 3 galaxies, one of which is
known to be underluminous in Ly$\alpha$, likely due to dust
absorption (MG~1019+0534, Dey, Spinrad, \&
Dickinson 1995).  However, one could also imagine a Ly$\alpha$--radio
power dependency, similar to the known \o2 -- 1.4 GHz radio power
relation (McCarthy 1993), or that the MGs typically reside in dustier
or more cold--gas--rich host galaxies.

At longer wavelengths, where the spectra are less affected by small
number statistics, there are several significant differences between
the radio galaxy composites.  Most strikingly, the \Mg2 doublet is one
of the most prominent lines in the MG composite, with \Mg2 / \carb3
$\sim 1.2$ and Ly$\alpha$ / \Mg2 $\sim 4.4.$  In the 3C/MRC composite,
however, the \Mg2 doublet is relatively weak, with \Mg2 / \carb3 $\sim
0.4$ and Ly$\alpha$ / \Mg2 $\sim 41.$  The \He2 / \Mg2 line ratio also
shows significant variation between the composite radio galaxy
spectra:  \He2 / \Mg2 $\sim 0.8$ for the MG sample, while \He2 / \Mg2
$\sim 4.2$ for the 3C/MRC composite. These strong discrepancies are
difficult to interpret, though the MG line ratios are intriguingly more
similar to those of LINERs and quasars than to the 3C/MRC composite
radio galaxy or Seyfert IIs (see Table~\ref{compare}).  We note that
the \Mg2 line in the MG composite is composed from five spectra,
suggesting that small number statistics may still be the source
of the these suggestive trends.

The strengths of the various ionization stages of carbon are a more
useful diagnostic of the excitation mechanisms in active galaxies:
selecting emission lines from the same element avoids metallicity
dependencies, while the small wavelength separation of the lines
considered makes the conclusions relatively insensitive to reddening.
In Fig.~\ref{carbon} we plot \carbon4 / \carb3 vs. \carb3 / \car2 for
individual galaxies in the MG sample, the composite spectra from
Table~\ref{compare}, and the models described below.  We find that the
MG composite appears to be in a lower ionization state than the 3C/MRC
composite, with carbon line ratios more comparable to the LINER nucleus
of NGC~4579 than to the quasar composites.  Again, an alternative
explanation is that MGs typically reside in dustier galaxies than
3C/MRC sources.  Unfortunately, the limited optical window and high
redshift of the current sample deny us access to the line pairs
commonly used for dust extinction measurements, such as
H$\alpha$/H$\beta$, Ly$\alpha$/H$\beta$, and \He2/\he2b.  However, the
strength of the Ly$\alpha$ line, which is quite sensitive to dust,
implies that there is not an immense amount of reddening in the MG
composite, while the dramatic change in the \Mg2 strength between the
3C/MRC composite and the MG composite suggests that there is a
significant difference between the emission line regions in these two
radio galaxy populations.  Finally, the relatively high \car2 / \Ne4
ratio indicates that the MG composite is in a lower ionization state as
compared to the higher radio power 3C/MRC composite, despite the \Ne4
line residing redwards of the the carbon lines so that dust reddening
alone can not explain the line ratio difference.  This suggests that the
ionization state of radio galaxies correlates with radio power.  
In a detailed study of lower redshift ($z < 0.7$) southern 2~Jy
radio sources, however, Tadhunter \etal (1998) find 
ionization state, as measured by the \o2/[\ion{O}{3}]~$\lambda$5007 ratio,
does not correlate with radio power.

To test the hypothesis of a radio power---ionization state relation in
high redshift radio galaxies, we have amassed a sample of 30 radio
galaxies with published \carbon4 and \carb3 fluxes from the
literature.  In Fig.~5 we plot \carbon4 / \carb3 vs.  rest--frame 1.4
GHz radio power.  There is a slight tendency for the stronger radio
sources to have larger \carbon4 / \carb3 ratios, implying higher
ionization states.  To test the significance the correlation, we have
calculated the nonparametric Spearman rank correlation coefficient,
$\rho$, for this sample, and find a value of $\rho=0.340$.  Note that
this statistic is independent of cosmology.  The null hypothesis, that
no correlation exists between radio power and ionization state can be
marginally rejected:  the probability of obtaining a rank correlation
coefficient this high from a sample of 30 uncorrelated variables is
3.5\%.  If we restrict our analysis to the two larger data sets, \ie,
the Keck/MG and ultra-steep source (USS) samples, then $\rho = 0.445$
for a sample of 25 sources.  The probability of obtaining a rank
correlation coefficient this high from a sample of 25 uncorrelated
variables is 1.5\%, implying that the correlation is significant.  We
consider these statistical arguments suggestive, but not conclusive, of
a radio power---ionization state correlation in the UV spectra of
high--redshift radio galaxies.  Insufficient sources with both \He2 and
\Mg2 well--detected prevented a similar analysis to be done with the
\He2 / \Mg2 ratio.


We have computed simple, single--zone photoionization models using
CLOUDY (Ferland 1996).  The purpose of this exercise is not to
fabricate a definitive model of the radio galaxy spectrum, but rather
to demonstrate that a simple low density gas photoionization model can
reproduce the overall character of the spectrum and compare the
best--fit parameters with those derived for the 3C/MRC composite.
Following McL98, our calculations were made for an ionization bounded
slab of gas with a constant density of $n_e = 100$ cm$^{-3}$
illuminated by a power--law spectrum of ionizing radiation, $F_\nu
\propto \nu^{-\alpha}.$  We calculated models with a range of spectral
index, $\alpha,$ and ionization parameter, $U$.  For reference, McL98
found that $\alpha = 1.5$ and $\log U = -1.8$ provided the best fit to
the composite 3C/MRC spectrum.

We find that the carbon line diagnostics are not well--reproduced by
these simple models.  For example, the best--fit CLOUDY model of McL98
is represented by an asterisk in Fig.~4.  Though this model does a
reasonable job at reproducing the overall character of the 3C/MRC
emission line spectrum, the \carb3 / \car2 ratio is much higher in the
model than in the composite spectrum.  Fig.~2d of Allen \etal (1998) is
also enlightening:  for a given \carbon4 / \carb3 ratio, the \car2
emission line is again too strong in the comparison AGN spectra to be
fit by simple single--zone photoionization models.  Photoionization
models which incorporate both optically thick (ionization bounded)
clouds and optically thin (matter bounded) clouds, as described in
Binnette \etal (1996), are capable of reproducing the observed carbon
line diagnostics, as are the shock models of Dopita \& Sutherland
(1996), implying that (the MG) radio galaxies are more complicated than
our basic model.  Investigations also reveal that the high equivalent
width \Mg2 line is also difficult to reproduce with the simple model:
for $1.0 < \alpha < 1.5$ and $-1.0 < \log U < -2.0,$ no model is
capable of reproducing the \Mg2 / \carb3 ratio we find in the MG
composite radio galaxy spectrum, indicating that shocks and/or more
complicated photoionization scenarios are necessary to explain the
emission line spectra of these galaxies.

\section {Conclusions}

We present optical identifications, finding charts, and spectra for
seventeen new high--redshift radio sources selected from the MG 5~GHz
survey.  We select targets of moderately steep spectral index to
preferentially observe radio galaxies, and, indeed, fifteen of the
seventeen sources discussed herein are radio galaxies.  The remaining
two sources are moderately steep--spectrum, radio--loud quasars which
may be important in terms of unified models of extragalactic radio
sources.  The spectra were all taken at the W.M. Keck Telescopes and
are representative of the fainter MG identifications attempted thus
far, with typical $R$ magnitudes of $23 - 24.$

We construct a composite MG radio galaxy spectrum and compare it with
the higher--radio power composite 3C/MRC radio galaxy spectrum of
McL98.  We find that the MG radio galaxies typically exhibit lower
ionization state spectra than the 3C/MRC radio galaxies.  The \Mg2 emission
line is extremely strong in the MG composite relative to the other
rest--frame UV emission lines, with \Mg2 / \carb3 $\sim 1.5$ and \Mg2 /
\lya\, $\sim 0.3$.  Extensive modeling with single--zone photoionization
models are incapable of reproducing the high \Mg2 / \carb3 ratio,
indicating that shocks and/or more complicated photoionization
scenarios are producing the emission line spectra of these distant
radio galaxies.

We have amassed a large sample of high--redshift radio galaxies
with published \carbon4 and \carb3 line strengths.  Comparing the
\carbon4 / \carb3 ratio to the rest--frame 1.4~GHz radio power, we
find evidence for a correlation between ionization state and radio
power.  A likely interpretation is that the more powerful radio sources
are in an active phase when the central engine is emitting more
flux across the electromagnetic spectrum with the augmented UV flux
leading to higher ionization state spectra.  As we progress from the
strongest radio sources to weaker sources, we find the emission line
strengths attenuate, the ionization state of the emission line region
diminishes, and the stellar populations apparently become more
dominant.  This last effect is seen both in the diminished alignment
affect for weak radio sources and the discovery of several weak
radio sources at moderate redshift whose spectra are devoid of the
UV emission lines that dominate most radio galaxy spectra (\eg Spinrad
\etal 1997).  An alternative explanation is to invoke multiple
emission line regions whose relative contributions vary with radio
power. 

The high--redshift radio galaxies discussed herein are faint, and the radio
power--line strength correlation implies that long integrations with
the new generation of large aperture telescopes is necessary to measure
the emission line strengths and redshifts of these sources.  High--redshift
radio galaxies from a range of radio flux density will be
the key to further investigations of the ionization state--radio power
relation.

\acknowledgements

We are very grateful to B. Burke and Sam Conner at MIT for introducing
us to the MG sample and for supplying much of the astrometry for our
candidate subset.  We thank Aaron Barth, Wil van Breugel, and Pat
McCarthy for valuable discussion and useful comments on the
manuscript.  We acknowledge Chuck Steidel for graciously obtaining the
MDM images, Marc Davis and Steve Zepf for obtaining the spectra of
MG~0422+0816 and MG~0511+0143, Marc Davis and Jeffrey Newman for
obtaining the spectra of MG~0148+1028 and MG~0308+0720, and Pat
McCarthy for providing the composite 3C/MRC radio galaxy spectrum.  We
thank Brian McLeod who has imaged a subset of the MG sample in the K
band and T.  Bida, W. Wack and J. Aycock  for invaluable help during
our Keck runs.  
The authors also gratefully acknowledge the referee, Rogier Windhorst,
for useful comments.
DS acknowledges support from IGPP grant 99--AP026, AD
acknowledges the support of NASA grant HF--01089.01--97A and partial
support from a postdoctoral research fellowship at NOAO, HS
acknowledges support from NSF grant AST~95--28536.  DS, AD, and DS
acknowledge Peet for invaluable help at the observatories and for
facilitating coherent discussion.


\begin{deluxetable}{lclcc}
\tablewidth{0pt}
\tablecaption{Journal of the observations.}
\tablehead{
\colhead{Source}                          & 
\colhead{Observation}                     & \colhead{Observation}    &
\colhead{Telescope}                       & \colhead{Exposure} \nl
\colhead{} & \colhead{Type\tablenotemark{\dag}} & \colhead{Date} &
\colhead{} & \colhead{(seconds)}}
\startdata
MG~0018+0940 & I & 1995 Sep 01 & Keck & 600             \nl
             & S & 1995 Sep 01 & Keck & 1500            \nl
MG~0046+1102 & I & 1995 Sep 01 & Keck & 1050            \nl
             & S & 1995 Sep 01 & Keck & 3000            \nl
MG~0122+1923 & I & 1995 Sep 01 & Keck & 600             \nl
             & S & 1997 Sep 12 & Keck & 2400            \nl
MG~0148+1028 & I & 1996 Oct 15 & Lick & 1260            \nl
             & S & 1997 Dec 23 & Keck & 1800            \nl
MG~0308+0720 & I & 1996 Oct 15 & Lick & 600             \nl
             & S & 1997 Dec 23 & Keck & 1800            \nl
MG~0311+1532 & I & 1994 Jan 11 & MDM  & 600             \nl
             & S & 1995 Aug 31 & Keck & 3600            \nl
MG~0422+0816 & I & 1992 Nov 25 & Lick & 200             \nl
             & S & 1995 Oct 25 & Keck & 1800            \nl
MG~0511+0143 & I & 1992 Nov 25 & Lick & 1500            \nl
             & S & 1995 Oct 25 & Keck & 1800            \nl
MG~1142+1338 & I & 1994 Apr 07 & MDM  & 1800            \nl
             & S & 1995 Feb 03 & Keck & 2400            \nl
MG~1251+1104 & I & 1995 Mar 14 & Keck & 1200    \nl
             & S & 1995 Mar 15 & Keck & 4500            \nl
MG~1401+0921 & I & 1997 Jul 01 & Keck & 900             \nl
             & S & 1997 Jul 02 & Keck & 2400            \nl
MG~2037$-$0011 &I & 1994 Sep 30 & MDM  & 1500           \nl
             & S & 1995 Jul 23 & Keck & 3600            \nl
             & S & 1995 Jul 24 & Keck & 3600            \nl
             & S & 1995 Aug 31 & Keck & 3600            \nl
MG~2041+1854 & S & 1996 Jun 16 & Keck & 300             \nl
MG~2058+0542 & I & 1992 Aug 01 & Lick & 1400            \nl
             & I & 1993 Aug 17 & Lick & 1500            \nl
             & S & 1994 Jul 10 & Keck & 3000            \nl
MG~2109+0326 & I & 1993 Sep 14 & Lick & 1100            \nl
             & S & 1994 Jul 09 & Keck & 6550            \nl
MG~2121+1839 & I & 1992 Aug 01 & Lick & 1200            \nl
             & S & 1994 Jun 10 & Keck & 4500            \nl
MG~2308+0336 & I & 1995 Jul 25 & Keck & 600             \nl
             & S & 1995 Jul 23 & Keck & 2400    
\enddata
\tablenotetext{\dag}{I: imaging; S: spectroscopy.}
\label{obsns}
\end{deluxetable}

\begin{deluxetable}{lcll}
\tablenum{2}
\tablewidth{0pt}
\tablecaption{Astrometric data.}
\tablehead{
\colhead{Source}         & \colhead{}   & \colhead{$\alpha_{2000}$}      &
\colhead{$\delta_{2000}$}}
\startdata
MG~0018+0940    & R  & 00 18 55.23 & +09 40 06.9  \nl
                & O  & 00 18 55.24 & +09 40 06.8  \nl
                & A  & 00 18 53.93 & +09 40 24.6  \nl
MG~0046+1102    & R  & 00 46 41.40 & +11 02 52.6  \nl
                & O  & 00 46 41.38 & +11 02 52.5  \nl
                & A  & 00 46 43.88 & +11 02 33.3  \nl
MG~0122+1923    & R  & 01 22 29.95 & +19 23 39.1  \nl
                & O  & 01 22 29.90 & +19 23 38.6  \nl
                & A  & 01 22 31.27 & +19 24 10.2  \nl
MG~0148+1028    & R  & 01 48 28.85 & +10 28 21.3  \nl
                & O  & 01 48 28.83 & +10 28 22.0 \nl
                & A  & 01 48 26.23 & +10 27 52.3  \nl
MG~0308+0720    & R  & 03 08 41.90 & +07 20 44.3  \nl
                & O  & 03 08 41.98 & +07 20 44.9  \nl
                & A  & 03 08 40.35 & +07 21 18.3  \nl
MG~0311+1532    & R  & 03 11 56.89 & +15 32 54.8  \nl
                & O  & 03 11 56.83 & +15 32 55.4  \nl
                & A  & 03 11 54.52 & +15 32 49.7  \nl
MG~0422+0816    & R  & 04 22 24.00 & +08 16 19.2  \nl
                & O  & 04 22 23.97 & +08 16 18.7  \nl
                & A  & 04 22 23.95 & +08 16 31.2  \nl
MG~0511+0143    & R  & 05 11 04.77 & +01 41 57.8  \nl
                & O  & 05 11 04.76 & +01 42 00.3  \nl
                & A  & 05 11 02.89 & +01 41 54.7  \nl
MG~1142+1338    & R  & 11 42 23.6  &   +13 38 01.3  \nl
                & O  & 11 42 23.69 &   +13 38 01.4 \nl
                & A  & 11 42 22.80 &   +13 37 51.9 \nl
MG~1251+1104    & R  & 12 51 00.02 &   +11 04 19.9  \nl
                & O  & 12 51 00.02 &   +11 04 21.6  \nl
                & A  & 12 50 58.67 &   +11 04 45.7  \nl
MG~1401+0921    & R  & 14 01 18.3  &   +09 21 23.7  \nl
                & O  & 14 01 18.50 &   +09 21 21.2 \nl
                & A  & 14 01 16.66 &   +09 20 51.5 \nl
MG~2037$-$0011  & R  & 20 37 13.41 & $-$00 10 58.5  \nl
                & O  & 20 37 13.41 & $-$00 10 58.5  \nl
                & A  & 20 37 12.90 & $-$00 10 56.9  \nl
MG~2041+1854    & R  & 20 41 24.2  &   +18 55 02.0  \nl
                & O  & 20 41 24.09 &   +18 55 00.9 \nl
                & A  & 20 41 25.20 &   +18 55 11.1 \nl
MG~2058+0542    & R  & 20 58 28.95 &   +05 42 51.0  \nl
                & O  & 20 58 28.82 &   +05 42 50.7  \nl
                & A  & 20 58 30.13 &   +05 42 43.7  \nl
MG~2109+0326    & R  & 21 09 21.71 &   +03 26 52.7  \nl
                & O  & 21 09 21.80 &   +03 26 51.6 \nl
                & A  & 21 09 20.93 &   +03 26 30.8 \nl
MG~2121+1839    & R  & 21 21 25.48 &   +18 39 08.7  \nl
                & O  & 21 21 25.48 &   +18 39 09.0  \nl
                & A  & 21 21 25.35 &   +18 38 53.6  \nl
MG~2308+0336    & R  & 23 08 25.0 &    +03 37 03.0  \nl
                & O  & 23 08 25.15 &   +03 37 03.6 \nl
                & A  & 23 08 26.07 &   +03 36 22.3 \nl
\enddata
\tablecomments{R: radio source; O: optical counterpart; A: offset star.}
\label{astrom}
\end{deluxetable}

\begin{deluxetable}{lcccccccc}
\tablenum{3}
\tablewidth{0pt}
\tablecaption{Imaging Properties of the sample.}
\tablehead{
\colhead{Source}        & \colhead{z}   & \colhead{$R$}         & 
\colhead{$4 \times E_{\rm B-V}$}        &
\colhead{$S_{\rm 4.8 GHz}$}  & \colhead{$\alpha_{\rm 1.4 GHz}^{\rm 4.8 GHz}$}   &
\colhead{Size}          & \colhead{PA}  & 
\colhead{$\log L_{\rm 4.8 GHz}$} \nl
\colhead{} & \colhead{} & \colhead{(mag)} & \colhead{} & \colhead{(mJy)} &
\colhead{} & \colhead{(\arcsec)} & \colhead{(\deg)} &
\colhead{(erg s$^{-1}$ Hz$^{-1}$)}}
\startdata
MG~0018+0940 & 1.586 & 23.0 & 0.237 & 132 & 1.08 & 0.0 & \nodata & 34.70 \nl
MG~0046+1102 & 1.813 & 23.1 & 0.209 &  74 & 1.07 & 0.0 & \nodata & 34.61 \nl
MG~0122+1923 & 1.595 & 23.3 & 0.105 & 134 & 1.06 & 0.0 & \nodata & 34.70 \nl
MG~0148+1028 & 2.845 & 21.4 & 0.157 & 176 & 0.71 & 0.0 & \nodata & 35.39 \nl
MG~0308+0720 & 2.975 & 21.1 & 0.801 & 164 & 0.95 & 0.0 & \nodata & 35.56 \nl
MG~0311+1532 & 1.986 & 23.6 & 0.405 &  62 & 1.21 & 5.1 & 161     & 34.72 \nl
MG~0422+0816 & 0.294 & 20:\tablenotemark{\dag} & 0.721 &  68 & 1.06 & 0.0 & \nodata & 32.53 \nl
MG~0511+0143 & 0.596 & 22:\tablenotemark{\dag} & 0.393 &  98 & 1.06 & 3.9 & 171     & 33.4
2 \nl
MG~1019+0534 & 2.765 & 23.7 & 0.037 & 100 & 1.22 & 1.3 & 103 & 35.40 \nl
MG~1142+1338 & 1.279 & 23.9 & 0.093 & 149 & 1.03 & 0.0 & \nodata & 34.46 \nl
MG~1251+1104 & 2.322 & 24:  & 0.000 &  62 & 1.21 & 0.0 & \nodata & 34.94 \nl
MG~1401+0921 & 2.093 & 23.3 & 0.013 &  92 & 0.89 & 3.5 & 138     & 34.81 \nl
MG~2037$-$0011 & 1.512 & 24.8 & 0.397 & 119 & 1.03 & 0.0 & \nodata & 34.57 \nl
MG~2041+1854 & 3.056 & 20:\tablenotemark{\dag} & 0.441 &  217 & 0.76 & 0.0 & \nodata & 35.
10 \nl
MG~2058+0542 & 1.381 & 23.7 & 0.425 & 283 & 1.19:& 0.0 & \nodata & 34.90 \nl
MG~2109+0326 & 1.634 & 22.0 & 0.281 & 119 & 0.75 & 0.0 & \nodata & 34.55 \nl
MG~2121+1839 & 1.860 & 22.7 & 0.341 &  69 & 1.09 & 6.3 & 145     & 34.63 \nl
MG~2144+1928 & 3.592 & 23.5:& 0.449 &  58 & 1.54 & 8.5 & 177     & 35.76 \nl
MG~2308+0336 & 2.457 & 23:  & 0.173 & 148 & 0.85 & 3.0 & 175     & 35.20 \nl
\enddata
\tablenotetext{\dag}{estimated from non--photometric conditions}
\tablecomments{Keck magnitudes are in 2\arcsec\ apertures,
MDM magnitudes are in 2\farcs5 apertures, and Lick
magnitudes are in 4\arcsec\ apertures.  Keck and MDM images
are through an $R$ filter, while Lick observations are through
the Spinrad night--sky filter, $R_S$ (Djorgovski 1986).  We
assume $H_{0}=50 \kmsMpc,q_0=0$ to calculate the radio powers.  Uncertain
numbers are indicated with a colon.  Reddening corrections have only
been applied to the spectroscopy.}
\label{prop}
\end{deluxetable}

\begin{deluxetable}{lcccc}
\tablenum{4}
\tablewidth{0pt}
\tablecaption{4.85 GHz Flux Densities for the Sample}
\tablehead{
\colhead{Source}        &  \colhead{1986} &
\colhead{Becker \etal 1991\tablenotemark{\dag}}         & \colhead{Gregory \& Condon 1991}
 &
\colhead{Griffiths \etal 1995}}
\startdata
MG~0018+0940 & 132 & 156 & 159 $\pm$ 22 & 190 $\pm$ 14 \nl
MG~0046+1102 & 74 &  98 & 100 $\pm$ 15 & \nodata \nl
MG~0122+1923 & 134 & 115 & 117 $\pm$ 16 & \nodata \nl
MG~0148+1028 & 176 & 192 & 193 $\pm$ 27 & \nodata \nl
MG~0308+0720 & 164 & 161 & 165 $\pm$ 23 & 213 $\pm$ 15 \nl
MG~0311+1532 & 62 &  53 &  54 $\pm$  9 & \nodata \nl
MG~0422+0816 & 68 & 113 & 116 $\pm$ 17 & 102 $\pm$ 12 \nl
MG~0511+0143 & 98 & 104 & 107 $\pm$ 16 & 102 $\pm$ 12 \nl
MG~1019+0534 & 115 & 115 & 132 $\pm$ 19 & 100 $\pm$ 12 \nl
MG~1142+1338 & 149 & 125 & 127 $\pm$ 18 & \nodata \nl
MG~1251+1104 & 62 & \nodata        & \nodata      & \nodata \nl
MG~1401+0921 & 92 &  89 &  92 $\pm$ 14 &  66 $\pm$ 11 \nl
MG~2037$-$0011&119 & 114 & \nodata      & 179 $\pm$ 14 \nl
MG~2041+1854 & 217 & 173 & 178 $\pm$ 24 & \nodata \nl
MG~2058+0542 & 283 & 424 & 427 $\pm$ 59 & 356 $\pm$ 21 \nl 
MG~2109+0326 & 119 &  81 &  86 $\pm$ 14 &  75 $\pm$ 11 \nl
MG~2121+1839 & 69 &  61 &  65 $\pm$ 10 & \nodata \nl
MG~2144+1928 & 58 &  71 &  76 $\pm$ 11 & \nodata \nl
MG~2308+0336 & 148 & 158 & 163 $\pm$ 23 & 160 $\pm$ 13 \nl
\enddata
\tablenotetext{\dag}{Becker \etal 1991 report a 15\%\ error on all measurements.}
\tablecomments{All flux densities measured in mJy.  The 1986 flux densities for
MG~0511+0143 and MG~2308+0336 are from Bennett \etal (1986);  the
remaining 1986 flux densities are from Lawrence \etal (1986).}
\label{radflux}
\end{deluxetable}

\begin{deluxetable}{lccccc}
\tablewidth{0pt}
\tablenum{5}
\tablecaption{Observed emission lines.}
\tablehead{
\colhead{Source}                        & \colhead{Line ID}              &
\colhead{$\lambda_{\rm obs}$}    &
\colhead{$f\times10^{-17}$}                     & \colhead{W$_\lambda$}                  &
\colhead{z} \nl 
\colhead{} & \colhead{} & \colhead{(\AA)} & 
\colhead{(erg cm$^{-2}$ s$^{-1}$)}      & \colhead{(\AA)} & \colhead{}}
\startdata
MG~0018+0940    & \ion{C}{4} 1549  & 4004.4 &   8.1 &     145 & 1.584   \nl
                & \ion{He}{2} 1640 & 4243.7 &   4.2 &      92 & 1.588   \nl
                & \ion{C}{3}] 1909 & 4934.6 &   8.7 &     161 & 1.585   \nl
                & \ion{C}{2}] 2326 & 6016.7 &   6.5 &     137 & 1.587   \nl
                & [\ion{Ne}{4}] 3426 &6263.8&   3.1 &      90 & 1.584   \nl
                & \ion{Mg}{2} 2798 & 7236.7 &   8.8 &     227 & 1.585   \nl
                & \ion{He}{2} 3203 & 8291.3 &   3.1 &      64 & 1.589   \nl
MG~0046+1102    & \ion{C}{4} 1549 &  4357.2 &   6.5 &      99 & 1.812   \nl
                & \ion{He}{2} 1640 & 4615.8 &   5.5 &      83 & 1.814   \nl
                & \ion{C}{3}] 1909 & 5366.5 &   7.9 &     124 & 1.811   \nl
                & \ion{C}{2}] 2326 & 6542.6 &   7.4 &     120 & 1.813   \nl
                & [\ion{Ne}{4}] 3426 &6813.2&   1.9 &      29 & 1.811   \nl
                & \ion{O}{2} 2470 &  6947.5 &   2.0 &      30 & 1.813   \nl
                & \ion{Mg}{2} 2798 & 7875.6 &  11.8 &     176 & 1.814   \nl
MG~0122+1928    & \carbon4   & 4033.9 & 3.2 & 54 & 1.604 \nl
                & \He2       & 4262.9 & 3.8 & 81 & 1.599 \nl
                & \carb3     & 4958.0 & 3.2 & 60 & 1.597 \nl
                & \car2      & 6047.3 & 2.3 & 64 & 1.600 \nl
                & \Mg2       & 7273.3 & 3.1 & 77 & 1.598 \nl
MG~0148+1028    & Ly$\alpha$ &       4676.1 & 101.8 & 116 & 2.845 \nl
                & \carbon4   &       5965.3 & 116.3 & 128 & 2.851 \nl
                & \He2       &       6304.9 &  26.0 &  29 & 2.844 \nl
                & \oxy3      &       6400.8 &   8.4 &   9 & 2.849 \nl
                & \carb3     &       7333.8 &  57.8 &  67 & 2.842 \nl
MG~0308+0720    & Ly$\alpha$ &       4858.3 & 181.8 &  71 & 2.995 \nl
                & \N5        &       4905.8 & 529.3 & 221 & 2.956 \nl
                & \siliox    &       5587.0 &  56.3 &  26 & 2.991 \nl
                & \carbon4   &       6162.9 & 155.9 &  75 & 2.979 \nl
                & \carb3     &       7565.2 &  69.8 &  39 & 2.963 \nl
MG~0311+1532    & \carbon4   &       4634.1 &   3.4 & 160 & 1.991 \nl
                & \He2       &       4903.5 &   2.0 & 103 & 1.990 \nl
                & \carb3     &       5703.8 &   2.1 &  97 & 1.988 \nl
                & \car2      &      \nodata &\nodata&\nodata & \nodata \nl
                & \Ne4       &       7239.3 &   1.1 &  50 & 1.988 \nl
MG~1019+0534    & \lya       &       4584.3 &   8.4 & 268 & 2.770 \nl
                & \N5        &       4664.8 &   2.3 &  66 & 2.762 \nl
                & \carbon4   &       5838.9 &  10.4 & 257 & 2.769 \nl
                & \He2       &       6174.9 &   8.5 & 170 & 2.765 \nl
                & \carb3     &       7179.6 &   4.9 &  93 & 2.760 \nl
MG~1142+1338    & \carb3 & \nodata&  $<$3.0 & \nodata & \nodata \nl
                & \ion{C}{2}] 2326 & \nodata&  $<$1.6 & \nodata & \nodata \nl
                & [\ion{Ne}{5}] 3426 & 7783.6 &  5.4 &    112 & 1.272   \nl
                & [\ion{O}{2}] 3727 & 8495.1 &  13.6 &    232 & 1.279   \nl
MG~1251+1104    & Ly$\alpha$ & 4040.5 & 23.1 &     148 & 2.324   \nl
                & \ion{C}{4} 1549 & 5148.4 &  3.0 & \nodata & 2.324   \nl
                & \ion{He}{2} 1640 & 5453.2 &  3.0 & \nodata & 2.325   \nl
                & \ion{C}{3}] 1909 & 6326.7 &  5.2 &    298 & 2.314   \nl
                & \ion{C}{2}] 2326 & \nodata& $<$4.5 & \nodata & \nodata \nl
MG~1401+0921    & \ion{C}{4} 1549  & 4793.9 & 4.1 & 165 & 2.095 \nl
                & \ion{He}{2} 1640 & 5072.8 & 5.0 & 171 & 2.093 \nl
                & \ion{C}{3}] 1909 & 5901.1 & 3.4 &  83 & 2.091 \nl
                & \ion{C}{2}] 2326 & 7187.2 & 1.7 &  89 & 2.090 \nl
                & [\ion{Ne}{4}] 2423 & 7521.8 &2.9& 121 & 2.104 \nl
MG~2037$-$0011  & \ion{C}{3}] 1909 & 4782.2 & 4.4 & 123 & 1.505 \nl
                & \ion{C}{2}] 2326 & 5843.5 & 1.0 &  39 & 1.512 \nl
                & [\ion{Ne}{4}] 2424 & 6093.9 & 1.0 & 43 & 1.514 \nl
MG~2041+1854    & Ly$\alpha$ & 4986 & 788.5 &     141 & 3.100   \nl
                & \ion{Si}{4}] 1400 &5695:&130.8 &      28 & 3.05:   \nl
                & \ion{C}{4} 1549 & 6283 & 417.8 &      54 & 3.056   \nl
                & \ion{C}{3}] 1909 & 7785:& 68.3 &      25 & 3.08:   \nl
MG~2058+0542    & \ion{C}{3}] 1909 & 4541.3 &  7.1 &     152 & 1.379   \nl
                & \ion{C}{2}] 2326 & 5537.6 &  4.1 &     120 & 1.381   \nl
                & \ion{Mg}{2} 2798 & 6667.0 &  6.6 &     206 & 1.381   \nl
                & [\ion{Ne}{4}] 2423 & 5769.6 &  1.3 &    45 & 1.380   \nl
                & [\ion{Ne}{5}] 3426 & \nodata & $<$2.0 & \nodata & \nodata \nl
                & [\ion{O}{2}] 3727 & 8877.4 & 12.0 &     240 & 1.382   \nl
MG~2109+0326    & \ion{C}{4} 1549 & \nodata& $<$1.0 & \nodata & \nodata \nl
                & \ion{He}{2} 1640 & 4323.7 &  3.2 &     148 & 1.636   \nl
                & \ion{C}{3}] 1909 & 5029.4 &  2.1 &     105 & 1.635   \nl
                & \ion{C}{2}] 2326 & 6136.6 &  2.3 &     110 & 1.638   \nl
                & [\ion{Ne}{4}] 2423 & 6386.8 &1.3 &      40 & 1.635   \nl
                & \ion{Mg}{2} 2798 & 7364.9 &  0.9 &     22: & 1.631   \nl
MG~2121+1839    & \ion{C}{4} 1549  & 4431.7 &  5.3 &     134 & 1.861   \nl
                & \ion{He}{2} 1640 & 4709.3 &  1.4 &      28 & 1.872   \nl
                & \ion{C}{3}] 1909 & 5462.6 &  2.4 &      47 & 1.861   \nl
                & \ion{C}{2}] 2326 & 6675:  & 1.5: &     45: & 1.869   \nl
                & [\ion{Ne}{5}] 3426 & \nodata & $<$1.0 & \nodata & \nodata \nl
                & \ion{Mg}{2} 2798 & \nodata & $<$0.5 & \nodata & \nodata \nl
MG~2144+1928    & Ly$\alpha$ & 5586.8 & 61.5 &     686 & 3.596   \nl
                & \ion{C}{4} 1549 & 7112.2 &  5.8 &     234 & 3.589   \nl
                & \ion{He}{2} 1640& 7534.4 &  3.5 &     362 & 3.594   \nl
MG~2308+0336    & Ly$\alpha$ & 4209.2 & 29.3 &     284 & 2.462   \nl
                & \ion{N}{5} 1240 & 4288.2 &  5.7 &     118 & 2.458   \nl
                & \ion{C}{4} 1549 & 5357.9 &  6.3 &      91 & 2.458   \nl
                & \ion{He}{2} 1640 & 5669.1 & 3.9 &      78 & 2.457   \nl
                & \ion{C}{3}] 1909 & 6589.1 & 4.5 &     110 & 2.452   \nl
                & \ion{C}{2}] 2326 & 8036.9 & 8.3 &     412 & 2.455   \nl 
\enddata
\tablecomments{Uncertain measurements are indicated with a colon.  Values
for MG~1019+0534 are from Dey, Spinrad, \& Dickinson 1995.  Values for
MG~2144+1928 are from Maxfield \etal 1998.}
\label{lines}
\end{deluxetable}

\begin{deluxetable}{lccccc}
\tablewidth{0pt}
\tablenum{6}
\tablecaption{Composite Keck/MG radio galaxy spectrum.}
\tablehead{
\colhead{Line}  & \colhead{$\lambda$}   & \colhead{Flux Density}  &
\colhead{$W_\lambda$} & \colhead{FWHM}  & \colhead{Comment} \nl
\colhead{}      & \colhead{(\AA)}       & \colhead{} &
\colhead{(\AA)} & \colhead{(km s$^{-1}$)} & \colhead{}}
\startdata
Ly$\alpha$      & 1216 &  515 & 116 & 1130  & \nl
Ly$\alpha$\tablenotemark{\dag}      & 1216 &  606 & 106 & 1140  & \nl
\ion{N}{5}      & 1240 &   53 &  13 & 1420  & \nl
\ion{C}{4}      & 1549 &  131 &  31 & 1540  & doublet \nl
\ion{He}{2}     & 1640 &   94 &  23 & 1150  & \nl
\ion{C}{3}]     & 1909 &  100 &  28 & 1260  & \nl
\ion{C}{2}]     & 2326 &   83 &  30 & 1720  & \nl
[\ion{Ne}{4}]   & 2424 &   49 &  16 & 1780  & \nl
[\ion{O}{2}]    & 2470 &   17 &   5 &  930  & \nl
\ion{Mg}{2}     & 2800 &  116 &  42 & 2100  & doublet \nl
[\ion{O}{2}]    & 3727 &  188 & 142 & 1060  & \nl
\enddata
\tablenotetext{\dag}{Composite composed without MG~1019+0534, a dusty
radio galaxy know to be underluminous in \lya\, (Dey, Spinrad, \& Dickinson 1995).}
\tablecomments{All line strengths measured relative to \carb3, whose
flux has been arbitrarily scaled to 100.}
\label{composite}
\end{deluxetable}

\begin{deluxetable}{lccccccc}
\tablewidth{0pt}
\tablecaption{Comparisons with Other AGN and Models --- Normalized Line Flux Densities.}
\tablehead{
\colhead{Line}  & \colhead{$\lambda$}   & \colhead{HzRG}  &
\colhead{HzRG}  & \colhead{Seyfert II}  & \colhead{LINER} & 
\colhead{QSO}   & \colhead{QSO}         \nl
\colhead{}      & \colhead{(\AA)}       & \colhead{MG} &
\colhead{3C/MRC}& \colhead{}            & \colhead{NGC~4579} &
\colhead{Boyle} & \colhead{Francis}}
\startdata
Ly$\alpha$      & 1216 &  515\tablenotemark{\dag} & 1766 &    1000 &     239 &     485 &  
   253 \nl
\ion{N}{5}      & 1240 &   53 &   88 & \nodata & \nodata &     128 & \nodata \nl
\ion{C}{4}      & 1549 &  131 &  207 &     218 &      81 &     224 &     217 \nl
\ion{He}{2}     & 1640 &   94 &  181 &      37 &       4 &      26 &      62 \nl
\ion{C}{3}]     & 1909 &  100 &  100 &     100 &     100 &     100 &     100 \nl
\ion{C}{2}]     & 2326 &   83 &   52 & \nodata &      32 &      19 &      20 \nl
[\ion{Ne}{4}]   & 2424 &   49 &   51 & \nodata &    $<$8 & \nodata &       8 \nl
[\ion{O}{2}]    & 2470 &   17 &   23 & \nodata &       7 & \nodata & \nodata \nl
\ion{Mg}{2}     & 2800 &  116 &   43 &      33 &     113 &     113 &     117 \nl
[\ion{O}{2}]    & 3727 &  188 &  207 &     103 & \nodata & \nodata &     0.3 \nl
\enddata
\tablenotetext{\dag}{When composite is created without MG~1019+0534, a
dusty radio galaxy known to be underluminous in \lya (Dey, Spinrad, \& Dickinson 1995),
the value of 100 $\times$ \lya / \carb3 $= 606$.}
\tablecomments{All line strengths are measured relative to
\ion{C}{3}].  MG high--redshift radio galaxy (HzRG) composite derives from the current
work, 3C/MRC HzRG composite is from McL98, Seyfert II composite is from
Ferland \& Osterbrock (1986), UV spectrum of LINER nucleus of NGC~4579
is from Barth \etal (1996), and QSO composites are from Boyle (1990)
and Francis \etal(1991).}
\label{compare}
\end{deluxetable}

 

\begin{figure}
\figurenum{1}
\caption{Finding charts for the new identifications.
Fields are 1.5 arcmin square, with north at the top and east to the
left.  Identifications are indicated with two dashes.  The offset star
in each field is marked with a capital A.  Note that for MG~2308+0336 no
bright stars are within the field--of--view; instead an offset galaxy
is indicated.}
\label{plate}
\end{figure}

\newpage
\begin{figure}
\figurenum{1b}
\end{figure}

\newpage
\begin{figure}
\figurenum{1c}
\end{figure}

\begin{figure}
\figurenum{2}
\plotone{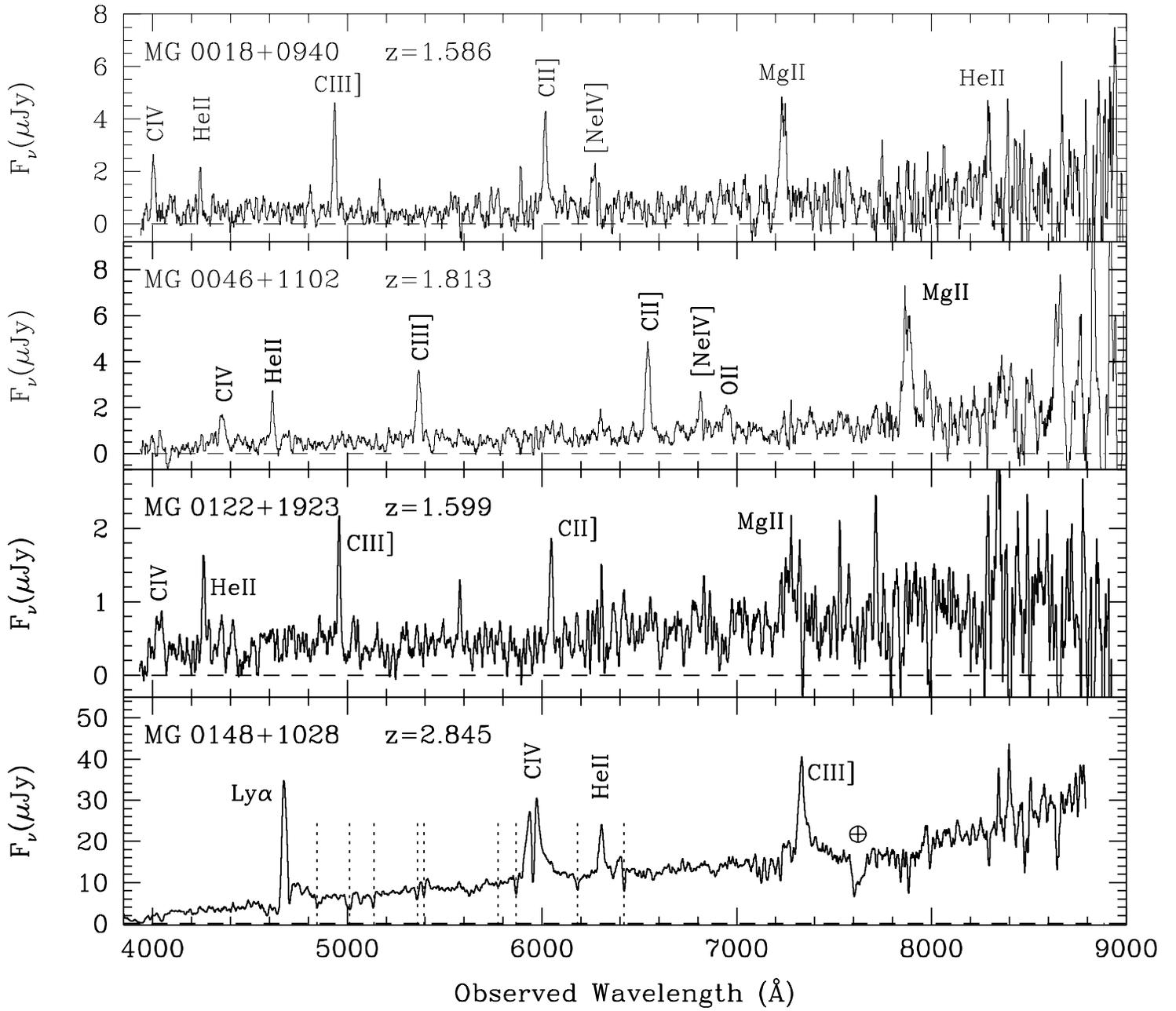}
\caption{\scriptsize{Spectra for the identifications, with prominent features
indicated.  Telluric absorption due to atmospheric water vaper (the
``A--band'') is indicated by a cross within a circle.  Vertical lines
in the spectrum of MG~0148+1028 are explained in the text.  The arrow
labeled ``D4000'' in the spectrum of MG~0511+0143 refers to the 4000
\AA\, break.  Spectra of MG~1019+0534 and MG~2144+1928 are from Dey, 
Spinrad, \& Dickinson (1995) and Maxfield \etal (1998) respectively.}}
\label{spectra}
\end{figure}

\begin{figure}
\figurenum{2b}
\plotone{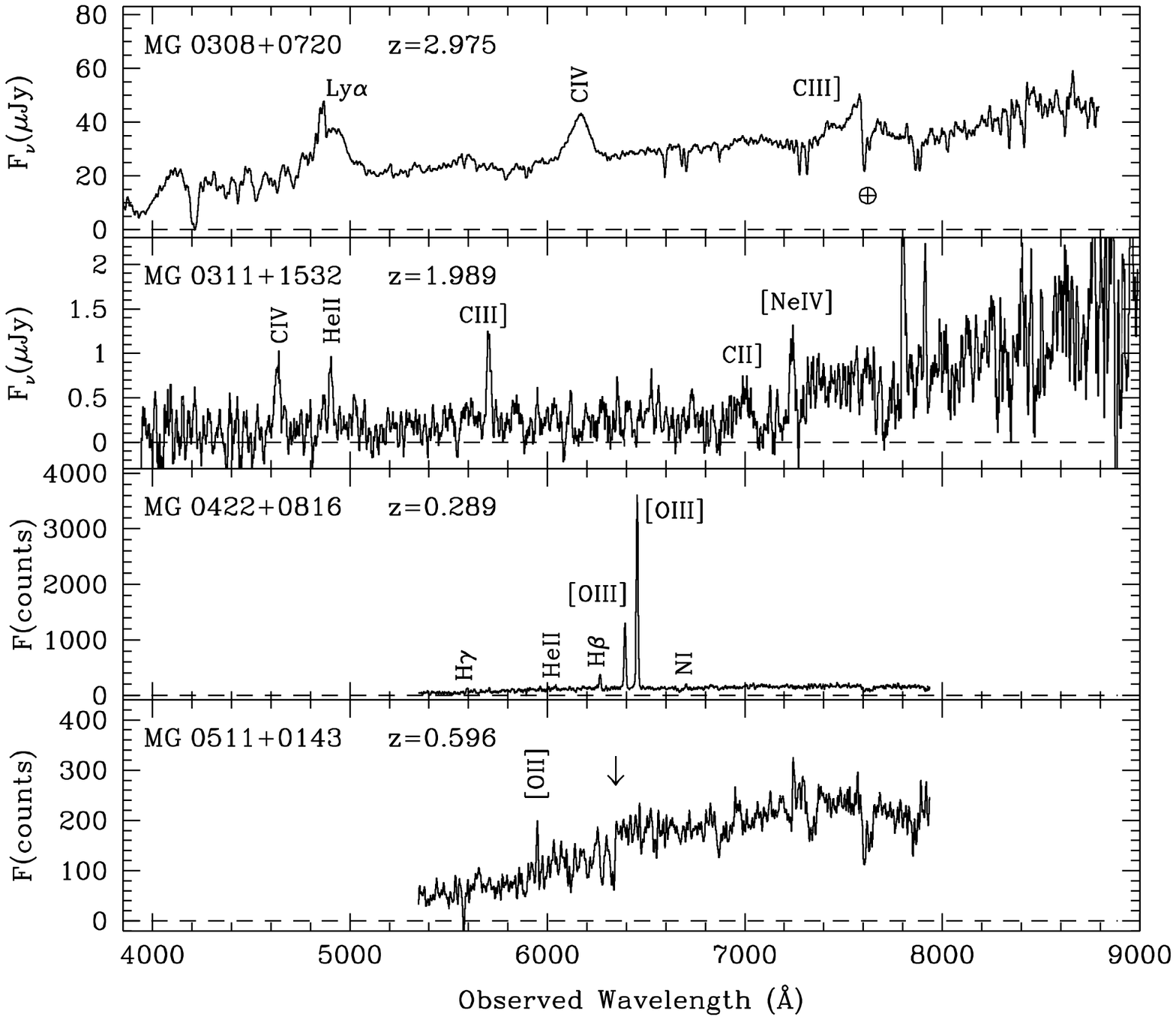}
\end{figure}

\begin{figure}
\figurenum{2c}
\plotone{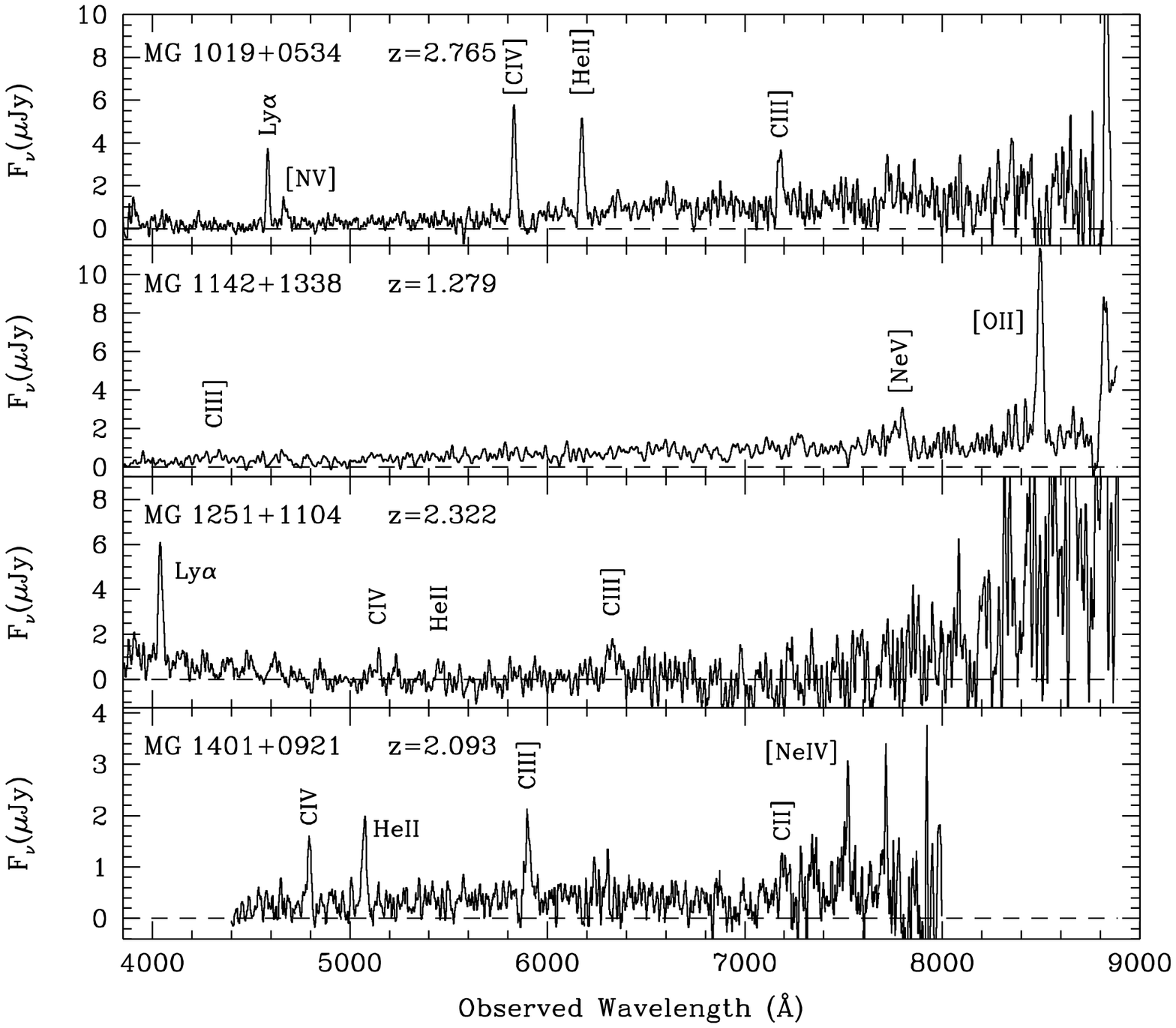}
\end{figure}

\begin{figure}
\figurenum{2d}
\plotone{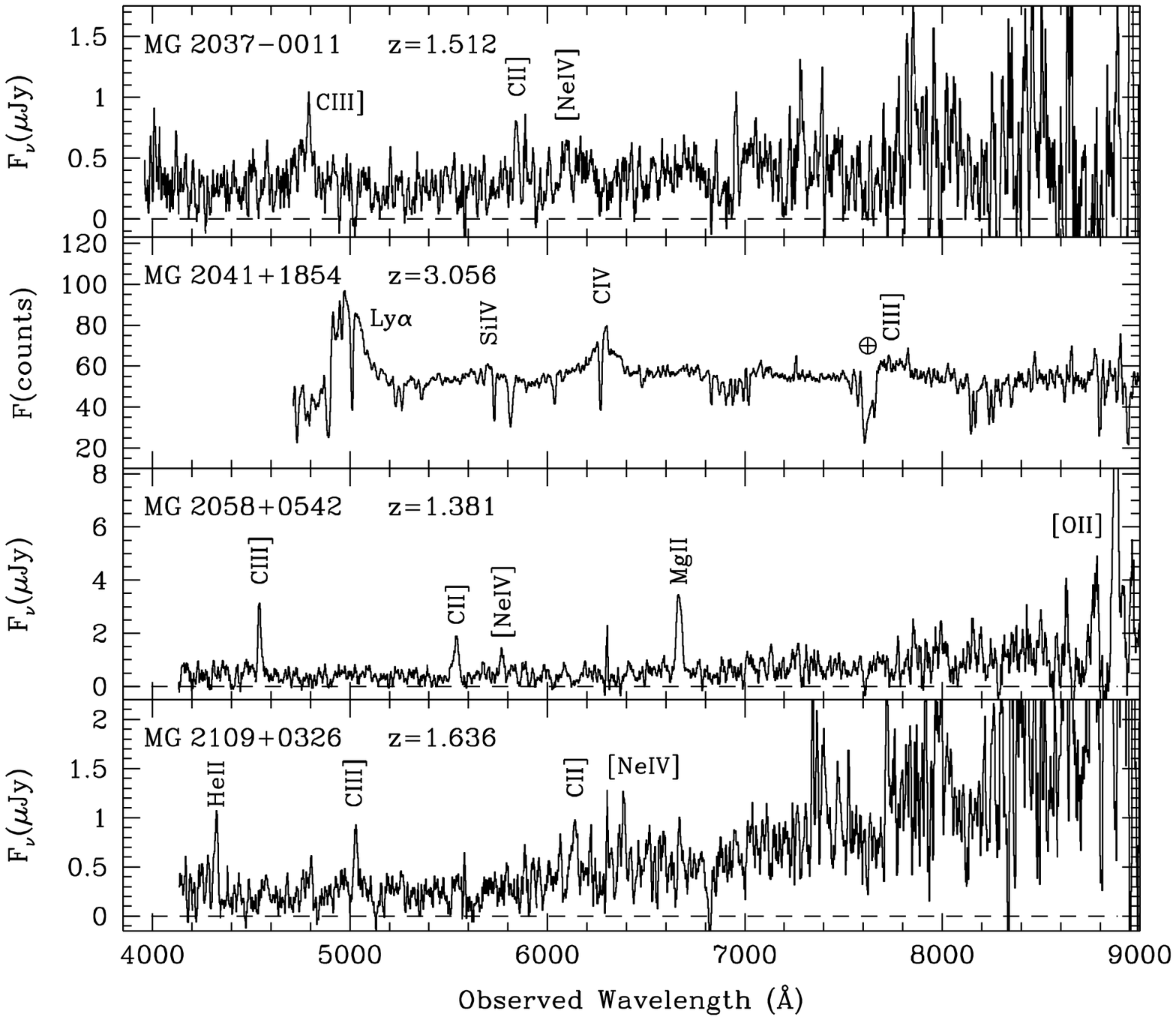}
\end{figure}

\begin{figure}
\figurenum{23}
\plotone{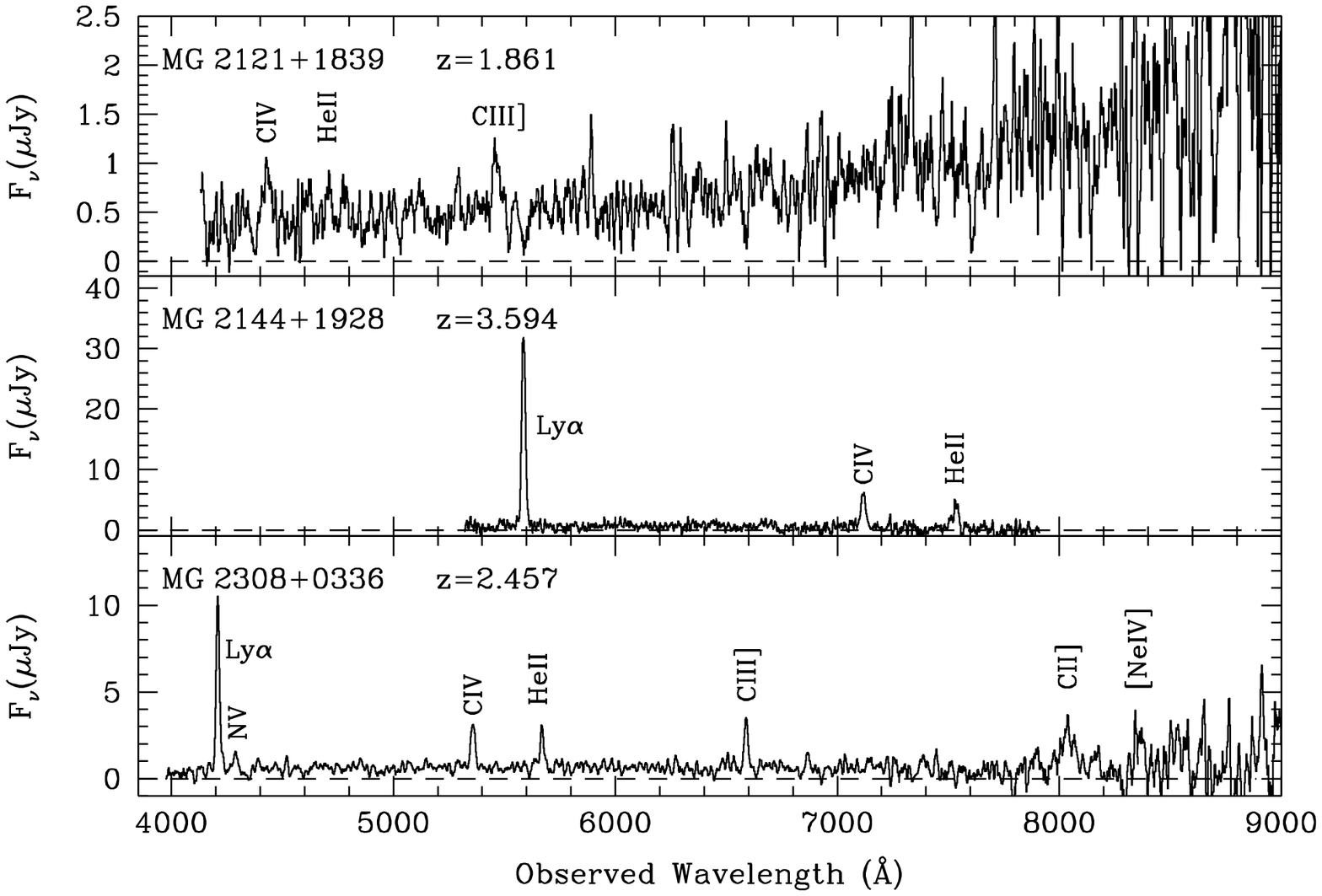}
\end{figure}

\begin{figure}[t]
\epsfxsize=5in
\epsffile{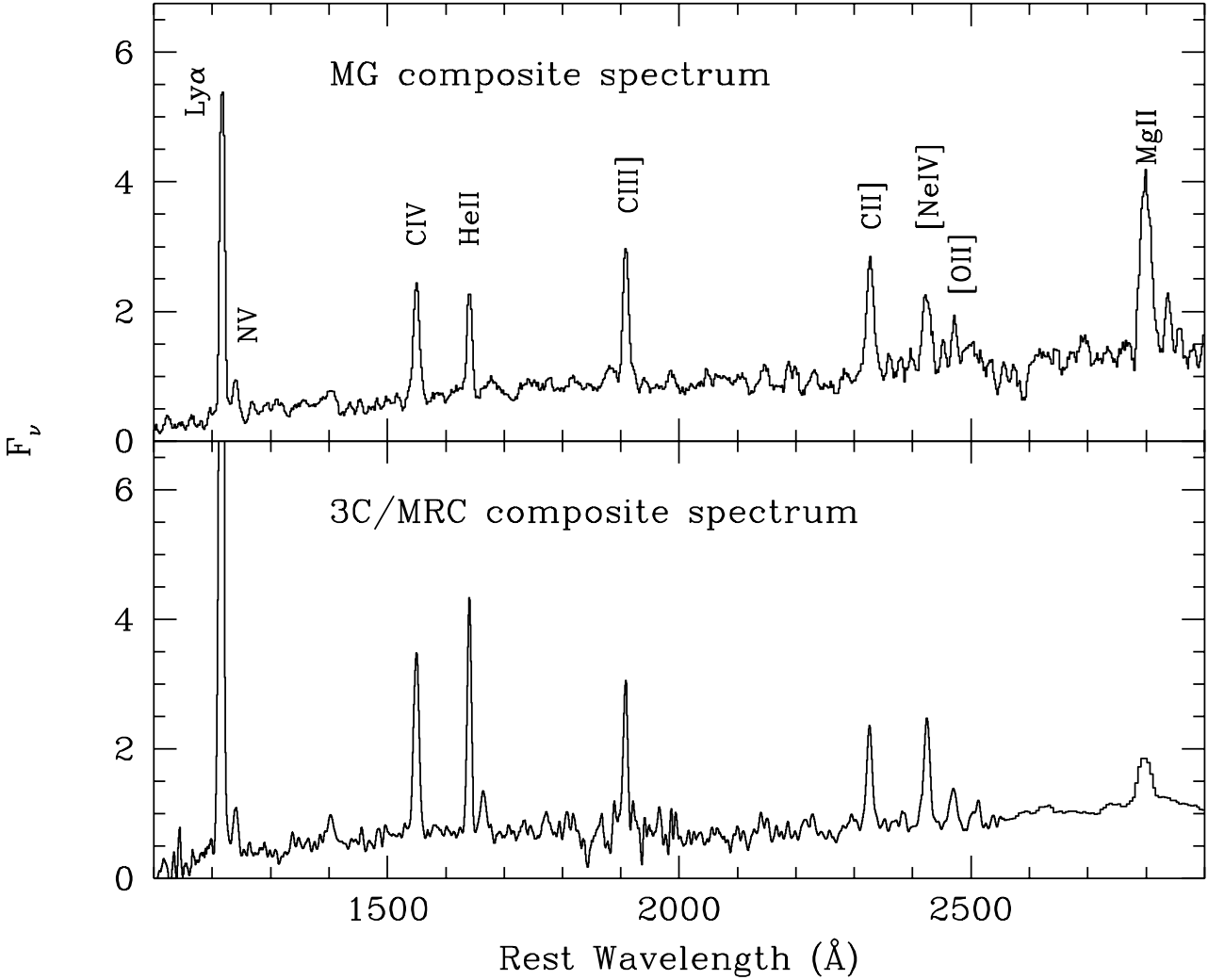}
\figurenum{3}
\caption{Composite MG (top) and 3C/MRC (bottom; McCarthy \& Lawrence
1997) radio galaxy spectra, scaled with respect to the \carb3 emission
line.  The 3C/MRC are typically higher radio flux density sources.
Note the difference in relative strength of the carbon lines: whereas
\carbon4 is the strongest of the carbon lines in the 3C/MRC composite,
it is the weakest of the carbon lines in the MG composite spectrum,
suggestive of a correlation between radio power and ionization state
in distant radio galaxies.}
\label{compfig}
\end{figure}

\begin{figure}[t]
\epsfxsize=6in
\epsffile{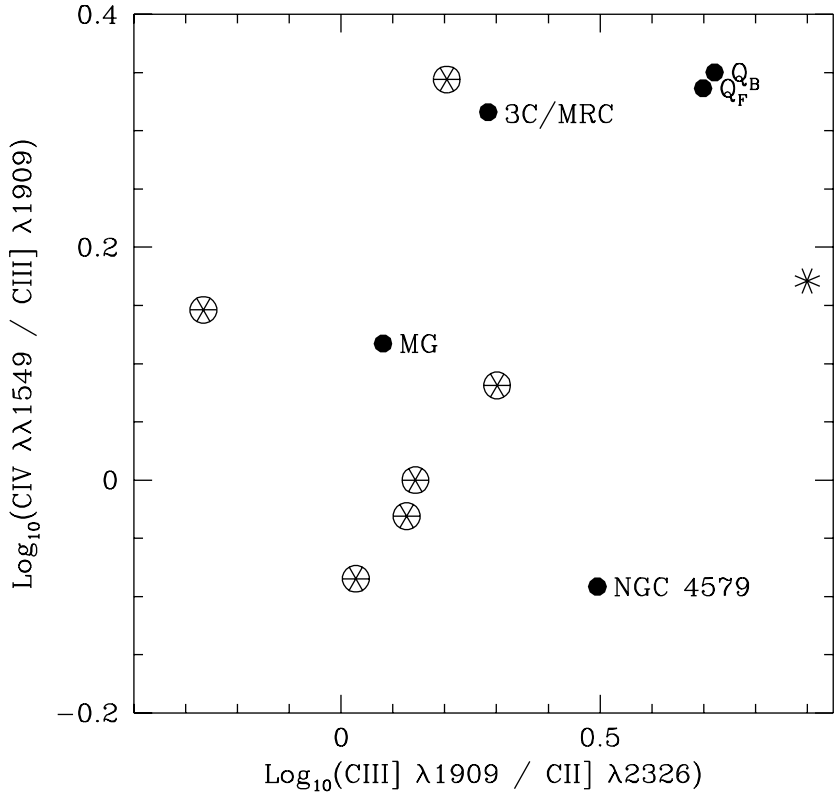}
\figurenum{4}
\caption{UV diagnostic diagram involving the various ionization stages
of carbon.  Pies represent individual MG galaxies.  Filled--in circles
with text refer to individual and composite spectra from
Table~\ref{compare}:  Q$_{\rm B}$ represents the quasar composite from
Boyle (1990); Q$_{\rm F}$ represents the quasar composite from Francis
\etal (1991); NGC~4579, a LINER galaxy, is from Barth \etal (1996).
Note that the MG galaxies and composite MG exhibit lower ionization
state carbon line ratios compared to the 3C/MRC composite and quasars.
Asterisk represents CLOUDY model (see text).} \label{carbon}
\end{figure}

\begin{figure}[t]
\epsfxsize=6in
\plotone{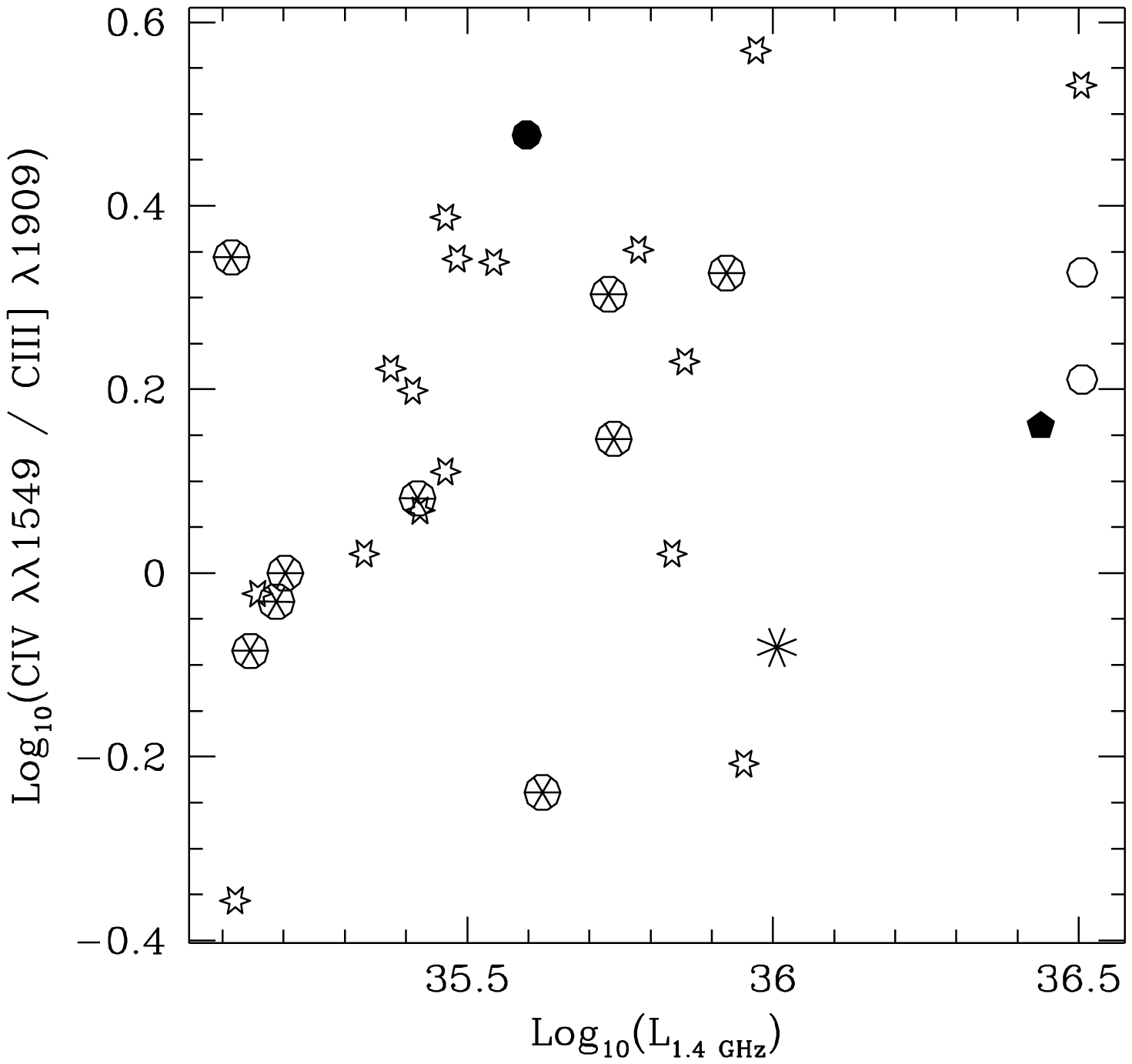}
\figurenum{5}

\caption{Ionization state--sensitive UV carbon line ratio plotted
against rest--frame 1.4 GHz radio power for $H_{0}=50 \kmsMpc$ and
$q_0=0$.  Pies represent individual MG galaxies, open stars are from
the ultra--steep sample (USS; R\"ottgering \etal 1994, van Ojik 1995),
asterisk represents 3C~294 (McCarthy \etal 1990), filled pentagon
represents 4C~41.17 (Dey \etal 1997), and filled/open circles represent
4C~00.54/4C~23.56 (Cimatti \etal 1998).}

\label{carpower}
\end{figure}


\begin{references}

\reference{alle98} Allen, M.G., Dopita, M.A., \& Tsvetanov, Zl. 1998, \apj, 493, 571
\reference{alli82} Allington--Smith, J.R. 1982, \mnras, 199, 611
\reference{armu97} Armus, L., Soifer, B.T., Murphy, T.W., Jr., Neugebauer, G., Evans, A.S., \& Matthews, K. 1997, \apj, 495, 276
\reference{bart96} Barth, A.J., Reichert, G.A., Filippenko, A.V., Ho, L.C., Shields, J.C., Mushotzky, R.F., \& Puchnarewicz, E.M. 1996, \aj, 112, 1829
\reference{beck91} Becker, R., Whiter, R., \& Edwards, A. 1991, \apjs, 75, 1
\reference{benn86} Bennett, C.L., Lawrence, C.R., Burke, B.F., Hewitt, J.N., \& Mahoney, J. 1986, \apjs, 61, 1
\reference{bine96} Binette, L, Wilson, A.S., \& Storchi--Bergmann, T. 1996,
\aa, 312, 365
\reference{blum79} Blumenthal, G., \& Miley, G.K. 1979, \aap, 80, 13
\reference{boyl90} Boyle, B.J. 1990, \mnras, 243, 231
\reference{brot94} Brotherton, M.S., Wills, B.J., Francis, P.J., \& Steidel, C.S. 1994, \apj, 430, 495
\reference{burn82} Burstein, D. \& Heiles, C. 1982, \aj, 87, 1165
\reference{card89} Cardelli, J.A., Clayton, C.C., \& Mathis, J.S. 1989, \apj, 345, 245
\reference{cham87} Chambers, R.C., Miley, G.K., \& van Breugel, W. 1987, \nat, 329, 604
\reference{cima98} Cimatti, A., di Serego Alighieri, S., Vernet, J., Cohen, M.
\& Fosbury, R.A.E. 1998, \apj, 499, 21
\reference{adey95} Dey, A., Spinrad, H., Dickinson, M. 1995, \apj, 440, 515 
\reference{adey97} Dey, A., van Breugel, W., Vacca, W., \& Antonucci, R. 1997, \apj, 490, 698
\reference{adey98} Dey, A., \etal 1998, in preparation
\reference{dick95} Dickinson, M., \etal 1996, in Fresh Views of Elliptical
Galaxies, ed. Buzzoni, Renzini, \& Serrano (ASP Conf. Ser., 86), 283
\reference{djor85} Djorgovski, S. 1985, \pasp, 97, 1119
\reference{dopi96} Dopita, M.A. \& Sutherland, R.S. 1996, ApJS, 102, 161
\reference{dunl93} Dunlop, J.S. \& Peacock, J. 1991, \mnras, 263, 936
\reference{dunl96} Dunlop, J.S., Peacock, J., Spinrad, H., Dey, A., Jimenez, R., Stern, D., \& Windhorst, R. 1996, Nature, 381, 581
\reference{eale93} Eales, S. \& Rawlings, S. 1993, \apj, 411, 67
\reference{ferl96} Ferland, G.J. 1996, Hazy:  A Brief Introduction to
CLOUDY, Univ. of Kentucky Dept. of Phys. and Astron. Internal Report
\reference{ferl86} Ferland, G.J. \& Osterbrock, D.E. 1986, \apj, 300, 658
\reference{fili82} Filippenko, A.V. 1982, \pasp, 94, 715
\reference{fran91} Francis, P., Hewett, P.C., Foltz, C.B., Chaffee, F.H., Weymann, R.J., \& Morris, S.L. 1991, \apj, 373, 465
\reference{greg91} Gregory, P.C. \& Condon, J.J. 1991, \apjs, 75, 1011
\reference{grif95} Griffiths, M., Wright, M., Burke, B., \& Ekers, R. 1995, \apjs, 97, 347
\reference{hale93} Hales, S.E.G., Baldwin, J.E., \& Warner, P.J. 1993, \mnras, 234, 919
\reference{knop97} Knopp, G.P. \& Chambers, K.C. 1997, \apjs, 109, 367
\reference{lacy94} Lacy, M., \etal 1994, \mnras, 271, 504
\reference{lawr86} Lawrence, C.R., Readhead, A.C.S., Moffet, A.T., \& Birkinshaw, M. 1986, \apjs, 61, 105
\reference{lowe97} Lowenthal, J. \etal 1997, \apj, 481, 673
\reference{mass88} Massey, P., Strobel, K., Barnes, J.V., \& Anderson, E.
1988, \apj, 328, 315
\reference{mass90} Massey, P. \& Gronwall, C. 1990, \apj, 358, 344
\reference{maxf98} Maxfield, L., \etal 1998, in preparation
\reference{mcca87} McCarthy, P., van Breugel, W., Spinrad, H., \& Djorgovski, S. 1987, \apj, 321, L29
\reference{mcca90} McCarthy, P., Spinrad, H., Dickinson, M., van Breugel, W., Liebert, J., Djorgovski, S., \& Eisenhardt, P. 1990, \apj, 365, 487
\reference{mcca93} McCarthy, P, 1993, \araa, 31, 639
\reference{mcca96} McCarthy, P., Kapahi, V.K., van Breugel, W., Persson, S.E., Athreya, R., \& Subrahmanya, C.R. 1996, \apjs, 107, 19
\reference{mcca98} McCarthy, P. \& Lawrence, C.R. 1998, in prepartion (McL98)
\reference{mill94} Miller, J.S. \& Stone, R.P.S. 1994, Lick Obs. Tech. Rep. 66
\reference{neus95} Neuschaefer, L.W. \& Windhorst, R.A. 1995, \apjs, 96, 371
\reference{oke95}  Oke, J.B., \etal 1995, \pasp, 107, 375
\reference{rawl91} Rawlings, S. \& Saunders, R. 1991, \nature, 349, 138
\reference{rawl96} Rawlings, S., Lacy, M., Blundell, K.\ M., Eales, S.\ A., Bunker, A.\ J., \& Garrington, S.\ T.\ 1996, \nature, 383, 502
\reference{huub94} R\"ottgering, H.J.A., Lacy, M., Miley, G.K., Chambers, K.C., \& Saunders, R. 1994, \aas, 108, 79
\reference{spin87} Spinrad, H. \& Djorgovski, S. 1987, in Observational Cosmology, ed. A. Hewitt, G. Burbridge, \& L.Z. Fang (IAU Symp. 124), 129
\reference{spin93} Spinrad, H., Dickinson, M., Schlegel, D., \& Gonzalez, R. 1993, in Observational Cosmology, ed. G. Chincarini, A. Iovino, \& D. Maccagni (ASP Conf. Ser., 51), 585
\reference{spin95} Spinrad, H., Dey, A., \& Graham, J.R. 1995, \apj, 438, L51
\reference{spin96} Spinrad, H., Dey, A., Stern, D., Dunlop, J., Peacock, J.,
Jimenez, R., \& Windhorst, R. 1997, \apj, 484, 581
\reference{spin98} Spinrad, H., Dey, A., Stern, \& Bunker, A. 1998, in Proc. Knaw Colloq., ed. H. R\"ottgering, in press
\reference{stei96} Steidel, C.C., Giavalisco, M., Pettini, M., Dickinson, M., \&
Adelberger K.L. 1996, \apj, 462, L17
\reference{ster96} Stern, D., Spinrad, H., Dickinson, M. 1996, \aj, 111, 102
\reference{ster97} Stern, D., Spinrad, H., Dey, A., Dickinson, M., \& Schlegel, D. 1997, in The HST and the
High--Redshift Universe, ed. N. Tanvir, A.  Aragon--Salamanca, \& J.
Wall (37th Herstmonceux), 413.
\reference{tadh98} Tadhunter, C.N., Morganti, R., Robinson, A., Dickson, R.,
Villar--Martin, M., \& Fosbury, R.A.E. 1998, \mnras, in press
\reference{thom94} Thompson, D., Djorgovski, S., Vigotti, M., \& Grueff, G. 1994 \aj, 108, 828
\reference{fool98} van Breugel, W., Stanford, S., Spinrad, H., Stern, D., \& Graham, J.R. 1998, \apj, 502, 614
\reference{ojik95} van Ojik, R. 1995, PhD thesis, Rijksuniversiteit te Leiden
\reference{vigo89} Vigotti, M., Grueff, G., Perley, R., Clark, B., \& Bridle,
A. 1989, \aj, 98, 419
\end{references}
\end{document}